\documentclass[12pt,a4paper]{article}
\pdfoutput=1
\usepackage{a4wide}
\usepackage{amsfonts}
\usepackage{amsmath}
\usepackage{amssymb}
\usepackage{array}
\usepackage{bigstrut}
\usepackage{booktabs}
\usepackage[small,bf]{caption}
\usepackage{cite}
\usepackage{color}
\usepackage{enumerate}
\usepackage{graphicx}
\usepackage{hyperref}
\usepackage{multirow}
\usepackage{pdflscape}

\def\1{\mathbf{1}}
\def\3{\mathbf{3}}
\def\2{\mathbf{2}}

\def\gtap{\ \raisebox{-.4ex}{\rlap{$\sim$}} \raisebox{.4ex}{$>$}\ }
\def\ltap{\ \raisebox{-.4ex}{\rlap{$\sim$}} \raisebox{.4ex}{$<$}\ }

\definecolor{darkgreen}{RGB}{34,181,34}
\definecolor{darkred}{RGB}{225,0,0}

\DeclareMathOperator{\diag}{diag}

\newcommand{\UPMNS}{U_\textrm{PMNS}}

\newcommand{\vev}[1]{ \left\langle {#1} \right\rangle }

\newcommand{\GEV}{ {\rm GeV} }

\def\diag{\mathop{\rm diag}\nolimits}

\def\UFN{U$(1)_\text{FN}$}
\def\QFN{Q_\text{FN}}

\numberwithin{equation}{section}

\begin{document}
\baselineskip 0.7cm
\begin{titlepage}
%%%%%%%%%%%%%%%%%%%%%%%

\vspace*{-15mm}
\begin{flushright} 
SISSA 60/2017/FISI  \\
IPMU17-0186
\end{flushright} 

\vskip 0.7cm
\begin{center}
{\bf \Large
Low-Scale Seesaw and the CP Violation \\
in Neutrino Oscillations
}
\vskip 1.2 cm

J.~T.~Penedo$^{~a}$, 
S.~T.~Petcov$^{~a,b,}$\footnote{Also at:
Institute of Nuclear Research and Nuclear Energy,
Bulgarian Academy of Sciences, 1784 Sofia, Bulgaria.} 
and
Tsutomu T.~Yanagida$^{~b,c}$ \\
\vspace{8mm}
$^{a}$\,{\it SISSA/INFN, Via Bonomea 265, 34136 Trieste, Italy} \\
$^{b}$\,{\it Kavli IPMU (WPI), University of Tokyo, 277-8583 Kashiwa, Japan} \\
$^{c}$\,{\it Hamamatsu Professor}

\vskip 1.2cm

\abstract{
We consider a version of the low-scale type I seesaw mechanism
for generating small neutrino masses,
as an alternative to the standard seesaw scenario.
It involves two right-handed (RH) neutrinos
$\nu_{1R}$ and $\nu_{2R}$ 
having a Majorana mass term with mass $M$,
which conserves the lepton charge
$L$. The RH neutrino $\nu_{2R}$ has 
lepton\discretionary{-}{-}{-}charge conserving
Yukawa couplings $g_{\ell 2}$ to the lepton and Higgs doublet fields,
while small lepton\discretionary{-}{-}{-}charge 
breaking effects are assumed to induce tiny lepton-charge
violating Yukawa couplings $g_{\ell 1}$ for $\nu_{1R}$, 
$l=e,\mu,\tau$. In this approach the smallness of neutrino masses 
is related to the smallness of the Yukawa coupling 
of  $\nu_{1R}$ and not to the large value of $M$: 
the RH neutrinos can have masses 
in the few GeV to a few TeV range. The Yukawa couplings  
$|g_{\ell 2}|$ can be much larger than  $|g_{\ell 1}|$, 
of the order $|g_{\ell 2}| \sim 10^{-4} - 10^{-2}$,
leading to interesting low-energy phenomenology.
We consider a specific realisation of this scenario
within the Froggatt-Nielsen approach to fermion masses.
In this model the Dirac CP violation phase $\delta$
is predicted to have approximately one of the values
$\delta \simeq \pi/4,\, 3\pi/4,$ or
$5\pi/4,\, 7\pi/4$,
or to lie in a narrow interval around one of 
these values. The low-energy phenomenology of 
the considered low-scale seesaw scenario of neutrino 
mass generation is also briefly discussed.
}

\end{center}
\end{titlepage}

\setcounter{footnote}{0}
\setcounter{page}{2}

%%%%%%%%%%%%%%%%%%%%%%% Introduction
\section{Introduction}
\label{sec:intro}
%%%%%%%%%%%%%%%%%%%%%%%
%
 The seesaw mechanism~\cite{seesaw} of neutrino mass generation 
is a very attractive mechanism which explains naturally 
the small masses of the neutrinos. 
According to the standard seesaw scenario 
the smallness of neutrino masses 
has its origin from large lepton-number
violating Majorana masses of right-handed (RH) neutrinos.
A very appealing aspect of the seesaw scenario is 
that we can relate the existence of large Majorana
masses of the RH neutrinos to a spontaneous breaking of 
some high scale symmetry, for example, GUT symmetry.
However, direct tests of the standard seesaw mechanism 
are almost impossible
due to the exceedingly large masses of the RH neutrinos.

 In the present article we consider  
an alternative mechanism for generating small neutrino masses. 
It involves two RH neutrinos $\nu_{1R}$ and $\nu_{2R}$ 
which have a Majorana mass $M\, \nu^{T}_{1R}\,C^{-1}\,\nu_{2R}$,
where $C$ is the charge conjugation matrix.
Assuming that $\nu_{1R}$ and $\nu_{2R}$ carry total lepton charges
$L(\nu_{1R}) = -1$  and $L(\nu_{2R}) =+1$, respectively,
this mass term conserves $L$. This implies that, as long as 
$L$ is conserved,  $\nu_{1R}$ and $\nu_{2R}$ (more precisely, $\nu_{1R}$ and 
$\nu_{2 L}^C \equiv C\,\overline{\nu_{2R}}^T$) form a heavy Dirac neutrino.
Since $L(\nu_{2R}) =+1$,  $\nu_{2R}$ can have lepton-charge 
conserving Yukawa couplings,
$-\mathcal{L} \supset g_{\ell 2}\,\overline{\nu_{2R}}\,H^{c\dagger}\, L_\ell$,
where $\ell=e,\mu,\tau$,  $L_\ell(x) = (\nu_{\ell L}(x)~~\ell_L(x))^T$ and 
$H^c = i \sigma_2 H^*$, $H = (H^+~~H^0)^T$ 
being the Higgs doublet field whose neutral component
acquires a vacuum expectation value (VEV).
On the other hand, the RH neutrino $\nu_{1R}$ cannot have a 
neutrino Yukawa coupling as long as lepton charge $L$ is conserved. 

 We assume further that some small lepton\discretionary{-}{-}{-}charge 
breaking effects induce tiny lepton\discretionary{-}{-}{-}charge violating Yukawa couplings
for $\nu_{1R}$, namely 
$-\mathcal{L} \supset g_{\ell 1}\,\overline{\nu_{1R}}\,H^{c\dagger}\, L_\ell$, 
$\ell=e,\mu,\tau$, with  $|g_{\ell 1}| \ll | g_{\ell' 2}|$,
$\ell,\ell' =e,\mu,\tau$. 
Our setup will imply that the lepton\discretionary{-}{-}{-}charge breaking
diagonal Majorana mass terms are either forbidden or suppressed.
In this case $\nu_{1R}$ and $\nu_{2R}$ 
(i.e., $\nu_{1R}$ and $\nu_{2 L}^C$) form a pseudo\discretionary{-}{-}{-}Dirac pair.
 In this scenario the smallness of neutrino masses 
is due to the small Yukawa coupling $|g_{\ell 1}| \ll 1$ and hence 
we do not have to introduce the large Majorana mass $M$ 
of the standard seesaw scenario. 
The mass $M$ of the $\nu^{T}_{1R}\,C^{-1}\,\nu_{2R}$  mass term 
can be at the weak scale. 

 The strong hierarchy  $|g_{\ell 1}| \ll | g_{\ell' 2}|$ 
between the two sets of Yukawa couplings  
can be realised rather naturally, for example, within 
the Froggatt-Nielsen (FN) scenario \cite{Froggatt:1978nt}.
Employing this scenario we will additionally consider
that the Yukawa couplings $g_{\ell 2}$
obey a standard FN hierarchy \cite{Sato:2000kj},
$|g_{e 2}| : |g_{\mu 2}| : |g_{\tau 2}| \,\sim\, \epsilon : 1 : 1$,
$\epsilon \sim 0.2$.
The magnitude of the Yukawa couplings of $\nu_{1R}$ should be 
completely different from that of the Yukawa coupling of $\nu_{2R}$.
However, due to the usual $\mathcal{O}(1)$ ambiguity in the
FN approach, it is impossible to predict unambiguously
the flavour dependence of $g_{\ell 1}$ and thus the
ratios $|g_{e 1}| : |g_{\mu 1}| : |g_{\tau 1}|$. 

 We show in the present article, in particular, that 
in the model of neutrino mass generation with two RH neutrinos 
with the hierarchy and flavour structure of their Yukawa couplings 
and the mass term outlined above
the Dirac CP-violating (CPV) phase is predicted to have 
one of the values 
$\delta \simeq \pi/4,\, 3\pi/4,$ or
$5\pi/4,\, 7\pi/4$.

%%%%%%%%%%%%%%%%%%%%%%% Setup
\section{General setup}
\label{sec:setup}
%%%%%%%%%%%%%%%%%%%%%%%

 We minimally extend the Standard Model (SM)
by adding two RH neutrinos,
i.e., two chiral fields $\nu_{aR}(x)$, $a = 1,2$,
which are singlets under the SM gauge symmetry group.
Following the notations of 
Refs.~\cite{Ibarra:2010xw,Ibarra:2011xn,Dinh:2012bp,Cely:2012bz},
the relevant low-energy Lagrangian is
%%%%%%%%%%%%%%%%%%%%%
\begin{equation}
\mathcal{L}_\nu\, =\, -\, \overline{\nu_{a R}} \, (M_D^T)_{a \ell}\,\nu_{\ell L}\,
-\frac{1}{2} \, \overline{\nu_{a R}} \, (M_N)_{a b}\, \nu_{b L}^C\,+\, \textrm{h.c.} \,, 
\label{eq:lagrangian}
\end{equation}
%%%%%%%%%%%%%%%%%%%%%
%
with $\nu_{a L}^C \equiv (\nu_{a R})^C  \equiv C\,\overline{\nu_{a R}}^T$,
$C$ being the charge conjugation matrix.
$M_N = (M_N)^T$ is the $2\times 2$ Majorana mass matrix of RH neutrinos,
while $M_D$ denotes the $3\times 2$ neutrino Dirac mass matrix,
generated from the Yukawa couplings of neutrinos following 
the breaking of electroweak (EW) symmetry. 
These Yukawa interactions read
%%%%%%%%%%%%%%%%%%%%%
\begin{equation}
\mathcal{L}_Y\, =\, -\, \overline{\nu_{a R}} \,(Y_D^T)_{a \ell }\,{H^c}^\dagger \,L_\ell
\,+\, \textrm{h.c.} \,,
\qquad M_D \,=\, v\,Y_D\,,
\label{eq:Yukawa}
\end{equation}
%%%%%%%%%%%%%%%%%%%%%
%
where $L_\ell(x) = (\nu_{\ell L}(x)~~\ell_L(x))^T$ and $H^c = i \sigma_2 H^*$,
$H = (H^+~~H^0)^T$ being the Higgs doublet field whose neutral component
acquires a VEV $v = \vev{H^0} =174$ GeV.
The matrix of neutrino Yukawa couplings has the form
%%%%%%%%%%%%%%%%%%%%%
\begin{equation}
Y_D \,\equiv\, 
\left(\begin{array}{cc}
g_{e 1}     & g_{e 2}   \\
g_{\mu 1}   & g_{\mu 2} \\
g_{\tau 1}  & g_{\tau 2} 
\end{array}\right)\,,
\label{eq:gdef}
\end{equation}
%%%%%%%%%%%%%%%%%%%%%
%
where $g_{\ell a}$ denotes the coupling of $L_\ell(x)$ to $\nu_{aR}(x)$, 
$\ell=e,\mu,\tau$, $a=1,2$.

 The full $5\times 5$ neutrino Dirac-Majorana mass matrix,
given below in the $(\nu_L,\nu_L^C)$ basis,
can be made block-diagonal by use of a unitary matrix $\Omega$,
%%%%%%%%%%%%%%%%%%%%%
\begin{equation}
\Omega^T
\left(\begin{array}{cc}
0     & M_D \\ 
M_D^T & M_N
\end{array}\right)
\,\Omega
\,=\,
\left(\begin{array}{cc}
U^* \hat{m} U^\dagger & 0 \\ 
0                     & V^* \hat{M} V^\dagger
\end{array}\right)\,,
\label{eq:block_diag}
\end{equation}
%%%%%%%%%%%%%%%%%%%%%
%
where $\hat{m} \equiv \diag(m_1,m_2,m_3)$ contains
the masses $m_i$ of the light Majorana neutrino mass eigenstates $\chi_i$,
while $\hat{M} \equiv \diag(M_1,M_2)$ contains
the masses $M_{1,2}$ of the heavy Majorana neutrinos, $N_{1,2}$.
Here, $U$ and $V$ are $3\times 3$ and $2 \times 2$
unitary matrices, respectively.
The matrix $\Omega$ can be parametrised as 
\cite{Ibarra:2010xw,Antusch:2009pm}:
%%%%%%%%%%%%%%%%%%%%%
\begin{equation}
\Omega \, = \,
\exp \left(\begin{array}{cc}
0     & R \\ 
-R^\dagger & 0 
\end{array}\right) \, = \,
 \left(\begin{array}{cc}
1 -\frac{1}{2}  R R^\dagger & R \\ 
-R^\dagger & 1 -\frac{1}{2} R^\dagger R
\end{array}\right)
+ \mathcal{O}(R^3)\,,
\label{eq:omega_par}
\end{equation}
%%%%%%%%%%%%%%%%%%%%%
%
under the assumption that the elements of the 
$3 \times 2$ complex matrix $R$ are small, which will be justified later. 
At leading order in $R$, the following relations hold~\cite{Ibarra:2010xw}:
%%%%%%%%%%%%%%%%%%%%%
\begin{align}
\label{eq:R}
R^*  \,&\simeq\, M_D\, M_N^{-1}\,, \\
\label{eq:mnu0}
m_\nu \,\equiv \,U^* \hat{m} U^\dagger \,&\simeq \,
R^* M_N R^\dagger - R^* M_D^T -M_D R^\dagger = -\, R^* M_N R^\dagger\,, \\
V^* \hat{M} V^\dagger \,&\simeq\,
M_N + \frac{1}{2} R^T R^* M_N+\frac{1}{2} M_N R^\dagger R \simeq\, M_N\,,
\label{eq:leading_R}
\end{align}
%%%%%%%%%%%%%%%%%%%%%
%
where
\footnote{The factors 1/2 in the two terms 
$\propto R^T R^* M_N$ and $\propto  M_N R^\dagger R$ 
in eq.~\eqref{eq:leading_R} are missing 
in the corresponding expression in Ref.~\cite{Ibarra:2010xw}.
These two terms provide a sub-leading correction to the leading term 
$M_N$ and have been neglected in the discussion of the phenomenology 
in Ref.~\cite{Ibarra:2010xw}. We will also neglect them 
in the phenomenological analysis we will perform.}
we have used eq.~\eqref{eq:R} to get the last equality 
in eq.~\eqref{eq:mnu0}. From 
the first two we recover the well-known seesaw formula
for the light neutrino mass matrix,
%%%%%%%%%%%%%%%%%%%%%
\begin{equation}
m_\nu \,=\, - M_D\, M_N^{-1}\,M_D^T \,.
\label{eq:seesaw}
\end{equation}
%%%%%%%%%%%%%%%%%%%%%
%

 We are interested in the case where only the $L$-conserving 
Majorana mass term of  $\nu_{1R}(x)$ and $\nu_{2R}(x)$, 
$M\, \nu^{T}_{1R}\,C^{-1}\,\nu_{2R}$, with  $M>0$ and, e.g.,  
$L(\nu_{1R}) = -1$  and $L(\nu_{2R}) =+1$, 
$L$ being the total lepton charge, is present in the Lagrangian.
In this  case the Majorana mass matrix of RH neutrinos 
$\nu_{1R}(x)$ and $\nu_{2R}(x)$ reads:
%%%%%%%%%%%%%%%%%%%%%
\begin{equation}
M_N \,=\, 
\left(\begin{array}{cc}
0 & M \\ M & 0
\end{array}\right)\,.
\label{eq:MN}
\end{equation}
%%%%%%%%%%%%%%%%%%%%%
%
Using eqs.~\eqref{eq:Yukawa}, \eqref{eq:gdef} and
eq.~\eqref{eq:seesaw}, we get the following expression 
for the light neutrino Majorana mass matrix $m_\nu$:
%%%%%%%%%%%%%%%%%%%%%
\begin{equation}
m_\nu \,=\,
- \frac{v^2}{M}
\left(\begin{array}{ccc}
2\,g_{e 1}\,g_{e 2} & g_{\mu 1}\,g_{e 2}+g_{e 1}\,g_{\mu 2} & 
g_{\tau 1}\,g_{e2}+g_{e1}\,g_{\tau 2} \\
g_{\mu 1}\,g_{e 2}+g_{e1}\,g_{\mu 2} & 2\,g_{\mu 1}\,g_{\mu 2} & 
g_{\tau 1}\,g_{\mu 2}+g_{\mu 1}\,g_{\tau 2} \\
g_{\tau 1}\,g_{e2}+g_{e1}\,g_{\tau 2} & g_{\tau 1}\,g_{\mu 2}+g_{\mu 1}\,g_{\tau 2} 
& 2\,g_{\tau 1}\,g_{\tau 2}
\end{array}\right)
\,.
\label{eq:mnu}
\end{equation}
%%%%%%%%%%%%%%%%%%%%%
%

 With the assignments $L(\nu_{1R}) = -1$  and $L(\nu_{2R}) =+1$ made, 
the requirement of conservation of the total lepton charge $L$  
leads to $g_{\ell 1} = 0$, $\ell= e, \mu, \tau$.
In this limit of $g_{\ell 1} = 0$, we have 
$m_\nu = 0$, the light neutrino masses vanish and 
$\nu_{1R}$ and $\nu_{2L}^C$ combine to form a Dirac fermion $N_D$
of mass 
$\tilde{M} \equiv \sqrt{M^2 + v^2\, \sum_\ell|g_{\ell 2}|^2}$~%
\footnote{\label{foot:L}%
These general results can be inferred 
just from the form of the conserved ``non-standard'' 
lepton charge $L^\prime$~\cite{Leung:1983ti}
which is expressed in terms of the individual lepton charges 
$L_\ell$, $\ell=e,\mu,\tau$, and 
$L_a(\nu_{bR}) = -\,\delta_{ab}$, $a,b=1,2$:
$L^\prime = L_e + L_\mu + L_\tau + L_1 - L_2$ 
($L^\prime(\nu_{1R}) = L_1(\nu_{1R}) = -1$  and 
$L^\prime(\nu_{2R}) = - L_2(\nu_{2R})  = +1$). 
Then ${\rm min}(n_+,n_-)$ and $|n_+ - n_-|$
are the numbers of massive Dirac and massless neutrinos, respectively, 
$n_+$ ($n_-$) being the number of charges entering into the expression 
for $L^\prime$ with positive (negative) sign.
},
%%%%%%%%%%%%%%%%%%%%%
\begin{equation}
N_D \,=\, 
\frac{N_1 \pm i\,N_2}{\sqrt{2}}  
\,=\,
\nu_{1R} + \nu_{2L}^C
\,,
\label{eq:ND}
\end{equation}
%%%%%%%%%%%%%%%%%%%%%
%
with $N_k = N_{kL} + N_{kR} \equiv N_{kL} + (N_{kL})^C 
= C\,\overline{N_k}^T$, $k=1,2$, 
and $\nu_{1R} = (N_{1R} \pm i\,N_{2R})/\sqrt{2}$,
$\nu_{2L}^C = (N_{1L} \pm i\,N_{2L})/\sqrt{2}$.

 Thus, the massive fields $N_k(x)$ are related to the 
fields $\nu_{aR}(x)$ by  
$\nu_{aR}(x) \simeq V_{ak}^*\, N_{kR}(x)$, where
%%%%%%%%%%%%%%%%%%%%%
\begin{equation}
V \,=\, 
\frac{1}{\sqrt{2}}
\left(\begin{array}{cc}
  1 & \mp i \\ 1 & \pm i
\end{array}\right)\,,
\label{eq:V}
\end{equation}
%%%%%%%%%%%%%%%%%%%%%
%
where the upper (lower) signs correspond to the case with 
the upper (lower) signs in eq.~\eqref{eq:ND} and in the expressions 
for $\nu_{1R}$ and $\nu_{2L}^C$ given after it.

 Small $L$-violating couplings $g_{\ell 1}\neq 0$ split 
the Dirac fermion $N_D$ into the two Majorana fermions $N_1$ and $N_2$ 
which have very close but different masses, $M_1 \neq M_2$, 
$|M_2 - M_1| \ll M_{1,2}$.
As a consequence, $N_D$ becomes a pseudo\discretionary{-}{-}{-}Dirac 
particle \cite{Wolfenstein:1981kw,Petcov:1982ya}. 
Of the three light massive neutrinos one remains massless 
(at tree level), while the other two acquire 
non\discretionary{-}{-}{-}zero and different masses.
The splitting between the masses of  $N_1$ and $N_2$
is of the order of one of the light neutrino mass differences 
and thus is extremely difficult to observe in practice.

 More specifically, in the case of a neutrino mass spectrum 
with normal ordering (see, e.g., \cite{Olive:2016xmw}) 
we have (at tree level) keeping terms up to 4th power 
in the Yukawa couplings $g_{\ell 1}$ and  $g_{\ell 2}$ 
and taking $g_{\ell a}$ to be real for simplicity:
%%%%%%%%%%%%%%%%%%%%%%%%%%%%%%%%%%%%%%%%%
\begin{equation}
 m_{1} \,=\, 0\,,\quad
m_{2,3} \,\simeq \,
\frac{1}{M}\left[
\sqrt{\Delta}  \left(1 - \frac{D(A^2 + \Delta)}{2M^2\Delta}\right)
\mp A          \left(1 - \frac{D}{M^2}\right)
\right]+\mathcal{O}(g_{\ell a}^6)\,,
\label{eq:lightmasses}
\end{equation}
%%%%%%%%%%%%%%%%%%%%%%%%%%%%%%%
%
where
%%%%%%%%%%%%%%%%%%%%%%%%%%%%%%%%%%%
\begin{align}
D \,&\equiv\, v^2\,\left(g^2_{e1} + g^2_{\mu 1}+ g^2_{\tau 1} + g^2_{e2} + g^2_{\mu 2} + g^2_{\tau 2}\right)\\
\Delta \,&\equiv\, v^4\,\left(g^2_{e1} + g^2_{\mu 1}+ g^2_{\tau 1} \right)
\left( g^2_{e2} + g^2_{\mu 2} + g^2_{\tau 2}\right)\,, \\
A \,&\equiv\,  v^2\left ( g_{e 1}\,g_{e2} + g_{\mu 1}\, g_{\mu 2}
+ g_{\tau 1} \, g_{\tau 2} \right )\,.
\end{align}
%
%%%%%%%%%%%%%%%%%%%%%%%%%%%%%%%
%
The heavy neutrino mass spectrum is given by:
%%%%%%%%%%%%%%%%%%%%%%%%%%%%%%%%%
\begin{equation}
  M_{1,2} \,\simeq\, 
   M \left[1 + \frac{D}{2M^2} - \frac{1}{2M^4} \left(\Delta + 2A^2 + \frac{D^2}{4}\right) \right]
  \mp \frac{A}{M}\left (1 - \frac{D}{M^2} \right )+\mathcal{O}(g_{\ell a}^6)\,.
\label{eq:heavymasses}
\end{equation}
%%%%%%%%%%%%%%%%%%%%%%%%%%%%%%%
%
The values of $m_{2,3}$ and $M_{1,2}$ given 
in eqs. (\ref{eq:lightmasses}) and (\ref{eq:heavymasses})
can be obtained as approximate 
solutions of the {\it exact} mass-eigenvalue equation:
%%%%%%%%%%%%%%%%%%%%%%%%%%%
\begin{equation}
\lambda^4 - \lambda^2\,\left ( M^2 + D \right ) 
- \,2\,\lambda\,M\,A -\, \left ( \Delta - A^2 \right ) = 0\,. 
\label{eq:masseigenveq}
\end{equation}
%%%%%%%%%%%%%%%%%%%%%%%%%%%%%%%
%

 Note that, as it follows from eqs. (\ref{eq:lightmasses}) and 
(\ref{eq:heavymasses}), we have \cite{Ibarra:2010xw}:
$M_2 - M_1 \simeq 2 (A/M)(1 - D/M^2) = m_3 - m_2$.
Therefore, the splitting between $M_2$ and $M_1$, 
as we have already noted, is exceedingly small. Indeed, 
for a neutrino mass spectrum with normal 
ordering (NO) and $m_1 = 0$, we have 
$m_2 = \sqrt{\Delta m^2_{21}} \simeq  8.6\times 10^{-3}$ eV, 
$m_3 = \sqrt{\Delta m^2_{31}} \simeq  0.051$ eV, and 
%%%%%%%%%%%%%%%%%%%%%%%%%%%%%%%
\begin{equation}
M_2 - M_1 \,=\, m_3 - m_2 \,\simeq\, 0.042~\textrm{eV}\,,
\label{eq:M2M1diff}
\end{equation}
%%%%%%%%%%%%%%%%%%%
%
where we have used the best fit
values of $\Delta m^2_{21}$ and $\Delta m^2_{31}$ determined 
in the recent global analysis of the neutrino oscillation data 
\cite{Capozzi:2017ipn}
(see also Table~\ref{tab:data}).
The corrections to the matrix $V$ which diagonalises 
$M_N$ are of the order of $AD/M^4$ and are negligible, 
as was noticed also in \cite{Ibarra:2010xw}.

To leading order in (real) $g_{\ell 1}$ and  $g_{\ell 2}$, the 
expressions in eqs.~\eqref{eq:lightmasses} and \eqref{eq:heavymasses} 
simplify significantly \cite{Ibarra:2010xw}:
%%%%%%%%%%%%%%%%%%%%%%%%%%%%%%%%
\begin{equation}
m_{1} \,=\, 0\,,\quad
 m_{2} \,\simeq \, \frac{1}{M}\left(\sqrt{\Delta} - A\right)\,,\quad
 m_{3} \,\simeq \, \frac{1}{M}\left(\sqrt{\Delta} + A\right)\,,
\label{eq:lightmassLO}
\end{equation}
%%%%%%%%%%%%%%%%%%%%%%%%%%%%%%%%
\begin{equation}
 M_{1} \,\simeq\, M \left(1 + \frac{D}{2M^2} \right) - \frac{A}{M}\,, \quad
 M_{2} \,\simeq\, M \left(1 + \frac{D}{2M^2} \right) + \frac{A}{M}\,.
\label{eq:heavymassLO}
\end{equation}
%%%%%%%%%%%%%%%%%%%%%%%%%%%%%%%
%

 The low-energy phenomenology involving the pseudo-Dirac 
neutrino $N_D$, or equivalently the Majorana neutrinos $N_1$ and $N_2$,
is controlled by the matrix  $RV$ of couplings of  $N_1$ and $N_2$ 
to the charged leptons in the weak charged lepton current
(see Section~\ref{sec:pheno}). 
When both $g_{\ell 1}$ and $g_{\ell 2}$ couplings are present,
this matrix is given by:
%%%%%%%%%%%%%%%%%%%%%
\begin{equation}
RV \,\simeq\, 
\frac{1}{\sqrt{2}}\,
\frac{v}{M}\,
\left(\begin{array}{cc}
g_{e 1}^*    + g_{e 2}^*     & i\,(g_{e 1}^*    - g_{e 2}^*) \\
g_{\mu 1}^*  + g_{\mu 2}^*   & i\,(g_{\mu 1}^*  - g_{\mu 2}^*)\\
g_{\tau 1}^* + g_{\tau 2}^*  & i\,(g_{\tau 1}^* - g_{\tau 2}^*) 
\end{array}\right)
\,,
\label{eq:RVgeneral}
\end{equation}
%%%%%%%%%%%%%%%%%%%%%
%
where we have used the expression for the matrix $V$ 
in eq.~\eqref{eq:V} with the upper signs.
We will adhere to this convention further on.

 It follows from the preceding discussion that 
the generation of non-zero light neutrino masses 
may be directly related to the generation of the $L$-non-conserving 
neutrino Yukawa couplings $g_{\ell 1}\neq 0$, $\ell=e,\mu,\tau$.
Among the many possible mechanisms leading 
to $g_{\ell 1}\neq 0$ there is at least one
we will discuss further,
that could lead to exceedingly small $g_{\ell 1}$, say  
$|g_{\ell 1}|\sim 10^{-12}-10^{-8}$.
In this case the RH neutrinos can have masses in 
the few GeV to a few TeV range and 
the neutrino Yukawa couplings  
$|g_{\ell 2}|$ can be much larger than  $|g_{\ell 1}|$, 
of the order $|g_{\ell 2}| \sim 10^{-4} - 10^{-2}$,
leading to interesting low-energy phenomenology.
For these ranges of $|g_{\ell 2}|$ and $M$, the approximations
$D/M^2 \ll 1$ and $\tilde{M}\simeq M$ are valid and will 
be used in what follows, i.e., we will use 
eqs.~\eqref{eq:lightmassLO} and \eqref{eq:heavymassLO}.

 Thus, in the scenario we are interested in
with two RH neutrinos possessing a Majorana mass 
term which conserves the total lepton charge $L$,  
the smallness of the light Majorana neutrino masses 
is related to the smallness of the $L$-non-conserving 
neutrino Yukawa couplings  $g_{\ell 1}$ and not to the 
RH neutrinos having large Majorana masses in the range of 
$\sim (10^{10} - 10^{14})$ GeV. 
Moreover, in contrast to the standard seesaw scenario,
the heavy Majorana neutrinos of the scenario of interest  
can have masses at the TeV or lower scale, which makes 
them directly observable, in principle, 
in collider (LHC, future $e^{+} - e^{-}$ and $p - p$) experiments.

 The low-scale type I seesaw scenario of interest 
with two RH neutrinos $\nu_{1R}$ and  $\nu_{2R}$
with $L$-conserving Majorana mass term and $L$-conserving
($L$-non-conserving) neutrino Yukawa couplings 
$g_{\ell 2}$ ($g_{\ell 1}$) of  $\nu_{2R}$ (of $\nu_{1R}$)  
was considered in \cite{Ibarra:2010xw} on purely 
phenomenological grounds
(see also, e.g., \cite{symmetry}).
It was pointed out in \cite{Ibarra:2010xw}, in particular, 
that the strong hierarchy $|g_{\ell 1}| \ll |g_{\ell' 2}|$, 
$\ell,\ell'=e,\mu,\tau$,
is a perfectly viable possibility from the point of view 
of generation of the light Majorana neutrino masses 
and that in this case the $L$\discretionary{-}{-}{-}non-conserving
effects would be hardly observable.
In the present article we provide a possible
theoretical justification of the strong hierarchy
between the $L$-conserving and  $L$-non-conserving 
neutrino Yukawa couplings based on the Froggatt-Nielsen 
approach to the flavour problem. 
We also investigate the phenomenology of this specific 
version of the low-scale type I seesaw model 
of neutrino mass generation, including 
the predictions for Dirac and Majorana leptonic 
CP violation.

%%%%%%%%%%%%%%%%%%%%%%% Model
\section{Froggatt-Nielsen Scenario}
\label{sec:model}
%%%%%%%%%%%%%%%%%%%%%%%

 We work in a supersymmetric (SUSY) framework
and consider a global broken
{\UFN} Froggatt\discretionary{-}{-}{-}Nielsen flavour symmetry,
whose charge assignments we motivate below.
We will show how an approximate U$(1)_\text{L}$ symmetry, 
related to the  $L$-conservation,
may arise in such a model, with $g_{\ell 1} \neq 0$ as
the leading $L$-breaking effect
responsible for neutrino masses.

 In our setup, one of the RH neutrino chiral superfields
has a negative charge under \UFN, namely $\QFN(\hat N_2) = -1$,
while the other carries a positive FN charge,
$\QFN(\hat N_1) \equiv n > 0$.
The FN mechanism is realised thanks to the VEV of the
lowest component $S$ of a chiral superfield $\hat S$,
which is a singlet under the SM gauge symmetry group
and carries negative FN charge, $\QFN(\hat S) = -1$.
Charges for the $\hat L_\ell$ superfields follow a standard
lopsided assignment \cite{Sato:2000kj},
namely $\QFN(\hat L_e) = 2$, $\QFN(\hat L_\mu) = 1$,
and $\QFN(\hat L_\tau) = 1$,
which allows for large $\nu_\mu$ -- $\nu_\tau$ mixing.
For definiteness we take $\QFN(\hat H_u) = 0$,
$\QFN(\hat e^c) = 4$,
$\QFN(\hat \mu^c) = 2$, and $\QFN(\hat \tau_c) = 0$.
The FN suppression parameter
$\epsilon \equiv \vev{S}/\Lambda$
is thus chosen to be close to the sine of the Cabibbo angle
$\lambda_C$, specifically $\epsilon = 0.2$,
in order to reproduce the hierarchies between charged lepton masses
(see also \cite{Kaneta:2017byo,Binetruy:1996cs}).
Here, $\Lambda$ is the FN flavour dynamics scale.
The charge assignments under \UFN~relevant to the present study
are summarised in Table \ref{tab:charges}.
%

%%%%%%%%%%%%%%%%%%%%%
\begin{table}
\centering
\renewcommand{\arraystretch}{1.2}
\begin{tabular}{ccccccccccc}
\toprule
  & $\hat S$ & $\hat N_1$ & $\hat N_2$ & $\hat H_u$ & $\hat L_e$ & $\hat L_\mu$ & $\hat L_\tau$ & $\hat e^c$ & $\hat \mu^c$ & $\hat \tau^c$ \\ 
\midrule
$\QFN$  & $-1$ & $n$ & $-1$ & 0 & 2 & 1 & 1 & 4 & 2 & 0 \\ 
\bottomrule
\end{tabular}
\caption{Charge assignments of lepton superfields under the \UFN~symmetry group.}
\label{tab:charges}
\end{table}
%%%%%%%%%%%%%%%%%%%%%

\pagebreak
 The effective superpotential~%
\footnote{The presence of an R-parity
preventing the usual $L$- and $B$-violating
terms in the MSSM superpotential is assumed.}
for the neutrino sector reads
%%%%%%%%%%%%%%%%%%%%%
\begin{equation}
W_\nu \,\sim\, 
M_0\,( \epsilon^{2n}\, \hat N_1\, \hat N_1
+ \epsilon^{n-1}\, \hat N_1\, \hat N_2)
+ (\epsilon\, \hat L_e + \hat L_\mu + \hat L_\tau)
 \, (\epsilon^{n+1}\,\hat N_1 + g_2\,\hat N_2) \, \hat H_u
\,,
\label{eq:W}
\end{equation}
%%%%%%%%%%%%%%%%%%%%%
where $M_0 \sim \Lambda$ and $g_2$ is
an {\it a priori} $\mathcal{O}(1)$ coupling.
Due to the condition of holomorphicity
of the superpotential, no quadratic term for $\hat N_2$
is allowed, justifying the absence of the Majorana
mass term $M\, \nu^{T}_{2R}\,C^{-1}\,\nu_{2R}$.
This framework may naturally arrange 
for the suppression $(M_N)_{11} \ll (M_N)_{12}$, as well as
for a hierarchy
between RH masses and the FN scale,
$M \sim \epsilon^{n-1}\, \Lambda \ll \Lambda$, 
provided the charge $n$ is sufficiently large.

 The limit of a large $\hat N_1$ charge, $n \gg 1$,
is quite interesting.
In this limit, one finds an accidental (approximate)
U$(1)_\text{L}$ symmetry, with
assignments $L(\hat N_{1,2}) = \pm 1$.
Furthermore, the desired hierarchy between 
(would-be) $L$-breaking and (would-be) $L$-conserving
Yukawa couplings, $|g_{\ell 1}| \sim \epsilon^{n+1} \ll |g_{\ell' 2}|$,
is manifestly achieved.
Finally, the mass term for $\hat N_1$ is
suppressed with respect to $\Lambda$
by the FN parameter to the power of $2n \gg 1$.
This observation and the 
holomorphicity of the superpotential
justify the absence of
diagonal Majorana mass terms $M\, \nu^{T}_{aR}\,C^{-1}\,\nu_{aR}$, $a=1,2$,
in eq.~\eqref{eq:MN} which could push up the light neutrino masses
to unwanted heavy scales.
We will focus on the case of a sufficiently large charge $n$
in what follows.

 The lopsided choice of FN charges
for the lepton doublets is responsible for the structure
$|g_{e 2}| : |g_{\mu 2}| : |g_{\tau 2}| \,\simeq\, \epsilon : 1 : 1$
of Yukawa couplings of $\nu_{2R}$.
However, due to the large FN charge of $\nu_{1R}$,
such FN flavour structure might be diluted in the
$L$-violating Yukawa couplings.
Indeed, for each insertion of $\hat S$, 
a factor of $\epsilon$ is in principle
accompanied by an $\mathcal{O}(1)$ factor.
This uncertainty makes it impossible to have an
unambiguous prediction for the ratios
$|g_{e 1}| : |g_{\mu 1}| : |g_{\tau 1}|$
in the model under discussion. This is
in contrast to the case of the $g_{\ell 2}$ couplings.
 
 Thus, in the present setup, the Yukawa matrix $Y_D$
obeys the following structure (up to phases):
%%%%%%%%%%%%%%%%%%%%%
\begin{equation}
Y_D 
\,\sim\, 
\left(\begin{array}{cc}
g_{e 1} & \epsilon \, g_2 \\
g_{\mu 1} & g_2 \\
g_{\tau 1} & g_2
\end{array}\right)\,
\sin \beta\,,
\label{eq:FNyukawas}
\end{equation}
%%%%%%%%%%%%%%%%%%%%%
%
with $\sin \beta = \vev{H_u^0}/v$,
and where $g_{\ell 1},g_2 > 0$,
and the hierarchy $g_{\ell 1} \ll g_2 \ltap 1$ is naturally realised.
We see from eq.~\eqref{eq:mnu} 
that the scale of light neutrino masses
depends on the size of the product $g_{\ell 1}\,g_2$,
namely
%%%%%%%%%%%%%%%%%%%%%%
\begin{equation}
\left( m_\nu \right)_{\ell \ell'} 
\,\sim\,
\frac{v^2\,\sin^2\beta}{M}
\, (g_{\ell 1} + g_{\ell' 1})\,g_2
\,.
\label{eq:magnitude}
\end{equation}
%%%%%%%%%%%%%%%%%%%%%

 Despite being suppressed, the quadratic term for $\hat{N}_1$,
 and thus the Majorana mass term $\mu\, \nu^T_{1R}\,C^{-1}\,\nu_{1R}$,
may still play a non-negligible role, for instance, in studies of leptogenesis \cite{Fukugita:1986hr}.
A complete suppression of $\mu$
can be achieved through the modification of our setup
which we summarise in the following.
Consider (4+1) dimensions where the extra dimension is compactified on
an $S^1/Z_2$ orbifold. This extra dimension has two fixed points, 
$y_1$ and $y_2$.
We localize all SM fields on $y_1$, 
a new chiral superfield $\hat \Phi$ (with lowest component $\Phi$) on $y_2$,
and allow the FN field $S$ and both RH neutrino fields to propagate in the bulk.
We impose, aside from the aforementioned FN symmetry ($\QFN(\Phi) = 0$), 
an U(1)$_{B-\hat{L}}$ symmetry with the charge assignments
$(B-\hat{L})(\nu_{1,2R}) = - 1$ and $(B-\hat{L})(\Phi) = +2$. 
Notice that $\hat{L}$ does not coincide with the standard (total) 
lepton charge $L$~\footnote{Indeed, 
we have  $\hat{L}(\nu_{1R}) = \hat{L}(\nu_{2R}) = + 1$
while $L(\nu_{1R}) = -L(\nu_{2R}) = -1$ (see Section \ref{sec:setup}).
}.
Then, interactions of the type 
$\Phi\, \nu^T_{aR}\,C^{-1}\,\nu_{bR}\,(a,b = 1,2)$
are allowed, provided a sufficient number of insertions of $S$ 
are considered. 
They generate mass terms for the RH neutrinos
once $\Phi$ develops a non-zero VEV, $\langle \Phi \rangle \neq 0$.
The Yukawa couplings $g_{\ell a}$ are allowed as before
and retain their FN hierarchy.
Assuming an enhanced U(1)$_{L}$ symmetry at $y_2$ 
with charges $L(\nu_{1R}) = -1$, $L(\nu_{2R}) = +1$ 
and $L(\Phi) = 0$, diagonal Majorana mass terms 
for $\nu_{1,2R}$ are thus forbidden.

%%%%%%%%%%%%%%%%%%%%%%% Mixing
\section{Neutrino Mixing}
\label{sec:mixing}
%%%%%%%%%%%%%%%%%%%%%%%

 The addition of the terms of eq.~\eqref{eq:lagrangian} to the SM Lagrangian
leads to a Pontecorvo-Maki\discretionary{-}{-}{-}Nakagawa-Sakata (PMNS) 
neutrino mixing matrix, $\UPMNS$, which is not unitary.
Indeed, the charged and neutral current weak interactions
involving the light Majorana neutrinos $\chi_i$ read:
%%%%%%%%%%%%%%%%%%%%%
\begin{align}
\mathcal{L}_\textrm{CC}^\nu \,&=\, - \frac{g}{\sqrt{2}}\,
\bar\ell\,\gamma_\alpha\,\big(U^\dagger_l (1+\eta) U\big)_{\ell i}\,\chi_{iL}\,W^\alpha
 \,+\, \textrm{h.c.} \,,\label{eq:CC0}\\
\mathcal{L}_\textrm{NC}^\nu \,&=\, - \frac{g}{2c_w}\,\overline{\chi_{iL}}\,\gamma_\alpha\,
\big(U^\dagger(1+2\eta) U\big)_{ij}\,\chi_{jL}\,Z^\alpha\,,
\label{eq:NC0}
\end{align}
%%%%%%%%%%%%%%%%%%%%%
%
where $\ell = e,\mu,\tau$ and $U_l$ 
is a unitary matrix which originates from the 
diagonalisation of the charged lepton mass matrix 
and $\eta \equiv -\, R\, R^\dagger/2$.
The transformation $U_l$
does not affect the power counting 
in the structure of eq.~\eqref{eq:FNyukawas},
though it may provide a source of deviations.
We then choose to work in the charged lepton mass basis,
in which $U_l = 1$. 
In this basis the PMNS neutrino mixing 
matrix is given by:
$\UPMNS  = (1+\eta) U$, where $U$ is the unitary matrix diagonalising 
the Majorana neutrino mass matrix generated by the seesaw mechanism
and $\eta$ describes the deviation from unitarity of the PMNS matrix.
As we will see further, the experimental constraints on the elements of 
$\eta$ imply $|\eta_{\ell \ell'}| \ltap 10^{-3}$, $\ell,\ell' = e,\mu,\tau$.

 Due to the structure of the matrix of Yukawa couplings $Y_D$
given in eq.~\eqref{eq:FNyukawas}, 
in the scheme we are considering the normal ordering (NO) of the
light neutrino mass spectrum, $m_1 < m_2 < m_3$, is favoured
over the spectrum with inverted ordering (IO), $m_3 < m_1 < m_2$.
We henceforth consider the NO case, 
for which, as we have already commented, 
we have $m_1 = 0$, $m_2 = \sqrt{\Delta m^2_{21}}$, and 
$m_3 = \sqrt{\Delta m^2_{31}}$.
Working in the basis of diagonal charged lepton mass term 
and neglecting the deviations from unitarity, 
which are parametrised by $\eta$,
we identify the PMNS mixing matrix with the unitary matrix 
$U$ which diagonalises $m_\nu$, $\UPMNS \simeq U$.
Given that one neutrino is massless (at tree level), the 
neutrino mixing matrix $U$ can be parametrised as:
%%%%%%%%%%%%%%%%%%%%%
\begin{equation}
\UPMNS \,=\, \left(
\begin{array}{ccc}
 c_{12}c_{13} & s_{12}c_{13} & s_{13}e^{-i\delta}\\
 -s_{12}c_{23} -c_{12}s_{23}s_{13}e^{i\delta} &  c_{12}c_{23} -s_{12}s_{23}s_{13}e^{i\delta} & s_{23}c_{13} \\
  s_{12}s_{23} -c_{12}c_{23}s_{13}e^{i\delta} & -c_{12}s_{23} -s_{12}c_{23}s_{13}e^{i\delta} & c_{23}c_{13}
\end{array}
\right)\,\diag(1,e^{i\alpha/2},1)\,,
\label{eq:UPMNS}
\end{equation}
%%%%%%%%%%%%%%%%%%%%%
%
where $c_{ij} \equiv \cos\theta_{ij}$ and 
$s_{ij} \equiv \sin\theta_{ij}$, with $\theta_{ij}\in[0,\pi/2]$,
while $\delta$ and $\alpha$ denote the Dirac and 
Majorana \cite{Bilenky:1980cx} CP violation (CPV) phases, 
respectively, $\delta, \alpha \in [0,2\pi]$.
The current best fit values and $3\sigma$ allowed ranges
for the neutrino mixing parameters and mass 
squared differences for NO spectrum 
are summarised in Table~\ref{tab:data}.
%%%%%%%%%%%%%%%%%%%%%
\begin{table}[t]
\centering
\renewcommand{\arraystretch}{1.2}
\begin{tabular}{lcc} 
\toprule
Parameter & Best fit value & $3\sigma$ range \\ 
\midrule
$\Delta m_{21}^{2}/10^{-5}$~eV$^2$ & $7.37$ & $6.93 - 7.96$\\
$\Delta m_{31}^{2}/10^{-3}$~eV$^2$ & $2.56$ & $2.45 - 2.69$\\
$\sin^2\theta_{12}/10^{-1}$ & $2.97$ & $2.50 - 3.54$ \\
$\sin^2\theta_{13}/10^{-2}$ & $2.15$ & $1.90 - 2.40$\\
$\sin^2\theta_{23}/10^{-1}$ & $4.25$  & $3.81 - 6.15$\\
$\delta/\pi$ & $1.38$ &
$[0,0.17] \cup [0.76,2]$\\
\bottomrule
\end{tabular}
\caption{Best fit values and 3$\sigma$ ranges of the 
neutrino oscillation parameters for neutrino mass spectrum 
with normal ordering (NO), obtained in the global analysis
of Ref.~\cite{Capozzi:2017ipn}.}
\label{tab:data}
\end{table}
%%%%%%%%%%%%%%%%%%%%%

%%%%%%%%%%%%%%%%%%%%%%% Phases
\section{Predictions for the CPV phases}
\label{sec:phases}
%%%%%%%%%%%%%%%%%%%%%%% 

 It proves convenient for our further analysis to use the 
Casas-Ibarra parametrisation~\cite{Casas:2001sr} 
of the Dirac mass matrix $M_{D}$ (neutrino Yukawa matrix $Y_{D}$):
%%%%%%%%%%%%%%%%
\begin{equation}
M_{D}\;=\;v\,Y_{D}\;=
\;i \,\UPMNS^{*}\,\sqrt{\hat{m}}\, O\, \sqrt{\hat{M}}\,V^{\dagger}\,,
\label{MDfromO}
\end{equation}
%%%%%%%%%%%%%%%%%
%
where $\hat{m}=\diag(m_{1},m_{2},m_{3})$ 
and $O$ is a complex orthogonal matrix. In the 
scheme with two heavy RH Majorana neutrinos 
the matrix $O$ has the form~\cite{Ibarra:2003up}:
%%%%%%%%%%%%%%%%%%%%%%%%%%%%%%%%
\begin{equation}
O\,\equiv\,\left(\begin{array}{cc}
				0 & 0\\
				\cos\hat{\theta} & \pm\sin\hat{\theta}\\
				-\sin\hat{\theta} & \pm\cos\hat{\theta}
\end{array}\right)\,, \quad \textrm{for NO mass spectrum,}
\label{eq:OforNO}
\end{equation}
%%%%%%%%%%%%%%%%%%%%%%%%%%%%%%%%
\begin{equation}
O\,\equiv\,\left(\begin{array}{cc}
				\cos\hat{\theta} & \pm\sin\hat{\theta}\\
				-\sin\hat{\theta} & \pm\cos\hat{\theta}\\
				0 & 0
\end{array}\right)\,, \quad \textrm{for IO mass spectrum,}
\end{equation}
%%%%%%%%%%%%%%%%%%%%%%%%%%%%%%%%
%
where $\hat{\theta}\equiv\omega-i\xi$.
The $O$-matrix in the case of  NO spectrum 
of interest can be 
decomposed as follows 
\footnote{A similar decomposition exists for 
the IO spectrum \cite{Ibarra:2011xn}.
}:
%%%%%%%%%%%%%%%%%%%%%%%%%%%%%%%%
\begin{equation}
O\,=\,
\frac{e^{i\hat\theta}}{2}
\left(
\begin{array}{cc}
0 & 0\\
1 & \mp i\\
i & \pm 1
\end{array}
\right) + 
\frac{e^{-i\hat\theta}}{2}
\left(
\begin{array}{cc}
0 & 0\\
1 & \pm i\\
-i & \pm 1
\end{array}
\right) 
\,=\, O_{+} + O_{-}\,.
\label{eq:Opm}
\end{equation}
%%%%%%%%%%%%%%%%%%%%%%%%%%%%%%%%%%
%
The Dirac neutrino mass matrix can be presented accordingly
as $M_D=M_{D+}+M_{D-}$, with obvious notation.
For the elements of $M_{D+} = v\,Y_{D+}$ and  $M_{D-} = v\,Y_{D-}$ we get:
%%%%%%%%%%%%%%%%%%%%%%%%%%%%%%%%
\begin{equation}
 v\,(Y_{D})_{\ell a} = v\,(Y_{D+})_{\ell a} + v\,(Y_{D-})_{\ell a} = 
v\,g^{(+)}_{\ell a} +  v\,g^{(-)}_{\ell a}\,,\quad\ell =e,\mu,\tau,~a=1,2\,, 
\label{eq:YDpm}
\end{equation}
%%%%%%%%%%%%%%%%%%%%%%%%%%%%%%%%%%%
%
where
%%%%%%%%%%%%%%%%%%%%%%%%%%%%%%%%%%
\begin{align}
\label{eq:gplusl1}
v\,g^{(+)}_{\ell 1} \,&\simeq\, 
i\,\frac{e^{i\omega} e^{\xi}}{2\sqrt{2}}\,
\left (\sqrt{M_1} \pm \sqrt{M_2} \right)\, 
\left (\sqrt{m_2}\,U^*_{\ell 2} + i\,\sqrt{m_3}\,U^*_{\ell 3}\right )\,,\\[3mm]
\label{eq:gplusl2}
v\,g^{(+)}_{\ell 2} \,&\simeq\, 
i\,\frac{e^{i\omega} e^{\xi}}{2\sqrt{2}}\,
\left (\sqrt{M_1} \mp \sqrt{M_2} \right)\, 
\left (\sqrt{m_2}\,U^*_{\ell 2} + i\,\sqrt{m_3}\,U^*_{\ell 3}\right )\,,\\[3mm]
\label{eq:gminl1}
v\,g^{(-)}_{\ell 1} \,&\simeq\, 
i\,\frac{e^{-i\omega} e^{-\xi}}{2\sqrt{2}}\,
\left (\sqrt{M_1} \mp \sqrt{M_2} \right)\, 
\left(\sqrt{m_2}\,U^*_{\ell 2} - i\,\sqrt{m_3}\,U^*_{\ell 3} \right )\,,\\[3mm]
\label{eq:gminl2}
v\,g^{(-)}_{\ell 2} \,&\simeq\, 
i\,\frac{e^{-i\omega} e^{-\xi}}{2\sqrt{2}}\,
\left (\sqrt{M_1} \pm \sqrt{M_2} \right)\, 
\left(\sqrt{m_2}\,U^*_{\ell 2} - i\,\sqrt{m_3}\,U^*_{\ell 3} \right )\,.
\end{align}
%%%%%%%%%%%%%%%%%%%%%%%%%%%%%%
%
Given the fact that $(\sqrt{M_2} - \sqrt{M_1})/(\sqrt{M_2} + \sqrt{M_1}) 
\simeq (m_3 - m_2)/(4M) \ll 1$ and, e.g., for $M = 10~(100)$ GeV, 
$(m_3 - m_2)/(4M) \simeq 10^{-12}~(10^{-13})$, 
it is clear from eqs.~\eqref{eq:gplusl1}~--~\eqref{eq:gminl2} that 
for  $\xi = 0$ we have (barring accidental cancellations):
$|g^{(-)}_{\ell 1}| \ll |g^{(+)}_{\ell'1}|$, 
$|g^{(+)}_{\ell 2}| \ll |g^{(-)}_{\ell'2}|$, 
$|g^{(+)}_{\ell 1}| \sim  |g^{(-)}_{\ell'2}|$, 
and thus $|g_{\ell 1}| \sim  |g_{\ell'2}|$, 
where we have used the upper signs in the expressions for 
$g^{(\pm)}_{\ell 1}$ and $g^{(\pm)}_{\ell 2}$. Unless otherwise stated 
we will employ this sign choice in the discussion which follows.
  
 Taking for definiteness $\xi< 0$, it follows 
from  eqs.~\eqref{eq:gplusl1}~--~\eqref{eq:gminl2} that
$|g^{(-)}_{\ell a}|$ ($|g^{(+)}_{\ell a}|$)
grows (decreases) exponentially
with $|\xi|$~
\footnote{Obviously, if $\xi > 0$, 
$|g^{(+)}_{\ell a}|$ ($|g^{(-)}_{\ell a}|$)
will grow (decrease) exponentially with $\xi$. 
}.
Therefore, for sufficiently large  $|\xi|$ 
we will have 
%%%%%%%%%%%%%%%%%%%%%%%%%%%%%%%%%
\begin{equation}
\frac{|g^{(+)}_{\ell 1}|}{|g^{(-)}_{\ell' 2}|} \,=\, 
e^{-2|\xi|}\, r_{\ell\ell'} \ll 1\,,
\quad
r_{\ell\ell'} \,\equiv\,
\frac{\left |\sqrt{m_2}\,U^*_{\ell 2} + i\,\sqrt{m_3}\,U^*_{\ell 3}\right |} 
{\left |\sqrt{m_2}\,U^*_{\ell' 2} - i\,\sqrt{m_3}\,U^*_{\ell'3}\right |}\,,
\quad
\ell,\ell'=e,\mu,\tau\,.
\label{eq:gplus1gminus2}
\end{equation}
%%%%%%%%%%%%%%%%%%%%%%%%%%%%%%%%%%
%
Using the $3\sigma$ allowed ranges of the neutrino 
oscillation parameters found in the global analysis of the 
neutrino oscillation data in \cite{Capozzi:2017ipn} and 
given in Table \ref{tab:data} 
and varying the CP violation phases in the PMNS matrix 
in their defining intervals it is not difficult to show that the 
ratios $r$ in eq.~\eqref{eq:gplus1gminus2}
vary in the interval $r_{\ell\ell'} = (0.04 - 22.5)$.

 Therefore even for the maximal cited value of $r_{\ell\ell'}$ 
we would have $|g^{(+)}_{\ell 1}| \ll |g^{(-)}_{\ell' 2}|$ 
for a sufficiently large value of $|\xi|$.
At the same time the inequalities
$|g^{(-)}_{\ell 1}|/|g^{(-)}_{\ell'2}| \ll 1$,
and $|g^{(+)}_{\ell 2}|/|g^{(-)}_{\ell'2}| \ll 1$, 
$\ell,\ell'=e,\mu,\tau$, always hold.
Thus, for $\xi < 0$ and sufficiently large  $|\xi|$ 
we get the requisite hierarchy of Yukawa couplings:
$|g_{\ell 1}|\simeq |g^{(+)}_{\ell 1}| \ll |g_{\ell'2}|\simeq |g^{(-)}_{\ell' 2}|$. 
For $|\xi| = 9$, for example, we find for $r_{\ell\ell'}\simeq 1$:
$|g_{\ell 1}|/ |g_{\ell '2}| \simeq 
|g^{(+)}_{\ell 1}|/|g^{(-)}_{\ell' 2}| \simeq 1.5\times 10^{-8}$,
which is in the range of values relevant for our discussion.
We get the same hierarchy of Yukawa couplings, $|g_{\ell 1}| \ll |g_{\ell'2}|$, 
$\ell,\ell'=e,\mu,\tau$, in the case of the lower signs in 
the expressions in eqs.~\eqref{eq:gplusl1}~--~\eqref{eq:gminl2}
for sufficiently large $\xi > 0$. In this case 
$|g_{\ell 1}|\simeq |g^{(-)}_{\ell 1}| \ll |g_{\ell'2}| \simeq |g^{(+)}_{\ell' 2}|$.

 We will show next that, given the present neutrino oscillation data,
enforcing the flavour pattern specified in eq.~\eqref{eq:FNyukawas}
results in a prediction for the Dirac phase $\delta$
close to $\pi/4$, $3\pi/4$, $5\pi/4$, $7\pi/4$,
and for the Majorana phase $\alpha$ close to zero.

 As we have seen, the matrix of neutrino Yukawa couplings 
$Y_D$ can be reconstructed up to normalization, 
a complex parameter, and a sign
using eqs.~\eqref{MDfromO} and \eqref{eq:Opm} 
(for NO spectrum).
For the cases of interest, with sufficiently large values of $|\xi|$, 
necessary to ensure the requisite hierarchy of Yukawa couplings
$|g_{\ell 1}|\ll |g_{\ell' 2}|$, $\ell,\ell'=e,\mu,\tau$,
the ratios of (absolute values of) Yukawa couplings read:
% 
%%%%%%%%%%%%%%%%%%%%%%%%%%%%%%%%
\begin{align}
R_{\ell\ell'}^{(1)} \,&\equiv\, \frac{|g_{\ell 1}|}{|g_{\ell' 1}|} 
\,\simeq\,
\frac{\left|\sqrt{m_2}\,U^*_{\ell 2} \pm i\,\sqrt{m_3}\,U^*_{\ell 3}\right|}{\left|\sqrt{m_2}\,U^*_{\ell' 2} \pm i\,\sqrt{m_3}\,U^*_{\ell' 3}\right|}
\,,\\[1mm]
R_{\ell\ell'}^{(2)} \,&\equiv\, \frac{|g_{\ell 2}|}{|g_{\ell' 2}|}
\,\simeq\,
\frac{\left|\sqrt{m_2}\,U^*_{\ell 2} \mp i\,\sqrt{m_3}\,U^*_{\ell 3}\right|}{\left|\sqrt{m_2}\,U^*_{\ell' 2} \mp i\,\sqrt{m_3}\,U^*_{\ell' 3}\right|}
\,,
\label{eq:ratios}
\end{align}
%%%%%%%%%%%%%%%%%%%%%%%%%%%%%%%%
%
where the upper and lower signs correspond to the case with $\xi < 0$
and upper signs in eq.~\eqref{eq:Opm}
and to the case with $\xi > 0$ and lower signs in eq.~\eqref{eq:Opm},
respectively.
Recall that
$|g_{\ell 1}| \simeq |g^{(+)}_{\ell 1}|$,
$|g_{\ell 2}| \simeq |g^{(-)}_{\ell 2}|$ 
in the former case ($\xi < 0$), and 
$|g_{\ell 1}| \simeq |g^{(-)}_{\ell 1}|$,
$|g_{\ell 2}| \simeq |g^{(+)}_{\ell 2}|$ 
in the latter ($\xi >  0$).

 One sees that the dependence on the
complex parameter $\hat\theta$ drops out
in the ratios $R^{(1,2)}_{\ell\ell'}$,
which are determined by the light neutrino masses
$m_2$ and $m_3$ and by neutrino mixing parameters only,
once the sign in $O$ in eq.~\eqref{eq:Opm}
(or equivalently in eqs.~\eqref{eq:gplusl1}~--~\eqref{eq:gminl2})
is fixed.
In particular, the flavour structure depends on
the elements $U_{\ell 2}$ and $U_{\ell 3}$ of
the PMNS matrix.
Given the fact that $m_2 = \sqrt{\Delta m^2_{21}}$, 
$m_3 = \sqrt{\Delta m^2_{31}}$,
and that $\Delta m^2_{21}$, $\Delta m^2_{31}$ and
the three neutrino mixing angles $\theta_{12}$, $\theta_{23}$ and $\theta_{13}$
have been determined in neutrino oscillation experiments with
a rather high precision, the quantities
$R_{\ell\ell'}^{(1)}$ and $R_{\ell\ell'}^{(2)}$ depend only
on the CPV phases $\delta$ and $\alpha$ once the sign of $\xi$ is fixed.
This means that knowing any two of the ratios $|g_{\ell 1}|/|g_{\ell' 1}|$ 
or $|g_{\ell 2}|/|g_{\ell' 2}|$, $\ell \neq \ell'=e,\mu,\tau$ allows to 
determine both $\delta$ and $\alpha$.

\begin{figure}[t]
\includegraphics[width=1.02\textwidth]{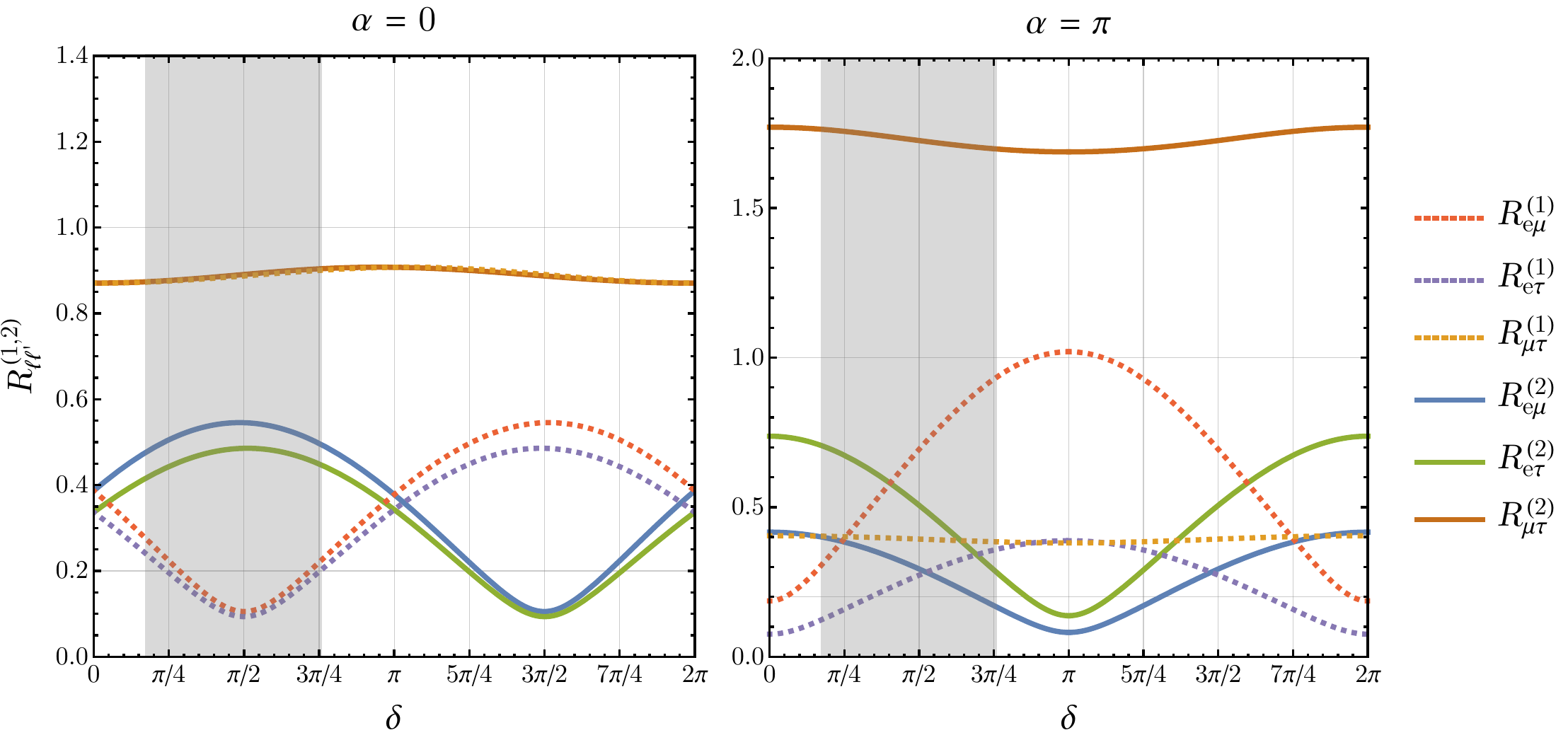}
\caption{Ratios $R^{(1,2)}_{\ell\ell'}$ of (absolute values of) Yukawa couplings
for a NO neutrino spectrum
as a function of the CPV phase $\delta$ for $\alpha = 0$ (left panel) and 
$\alpha = \pi$ (right panel),
in the case $\xi < 0$. The figure is obtained using the best fit values of 
$\Delta m^2_{21,31}$ and $\sin^2 \theta_{ij}$ quoted in Table \ref{tab:data}.
The vertical grey band indicates values of $\delta$ which are disfavoured at $3\sigma$.
The case $\xi > 0$ is obtained by exchanging $R^{(1)}_{\ell\ell'}$ and $R^{(2)}_{\ell\ell'}$. (For interpretation of the colours in the figure(s), the
reader is referred to the web version of this article.)
}
\label{fig:ratiosbfv}
\end{figure}

\begin{figure}[t]
\includegraphics[width=1.02\textwidth]{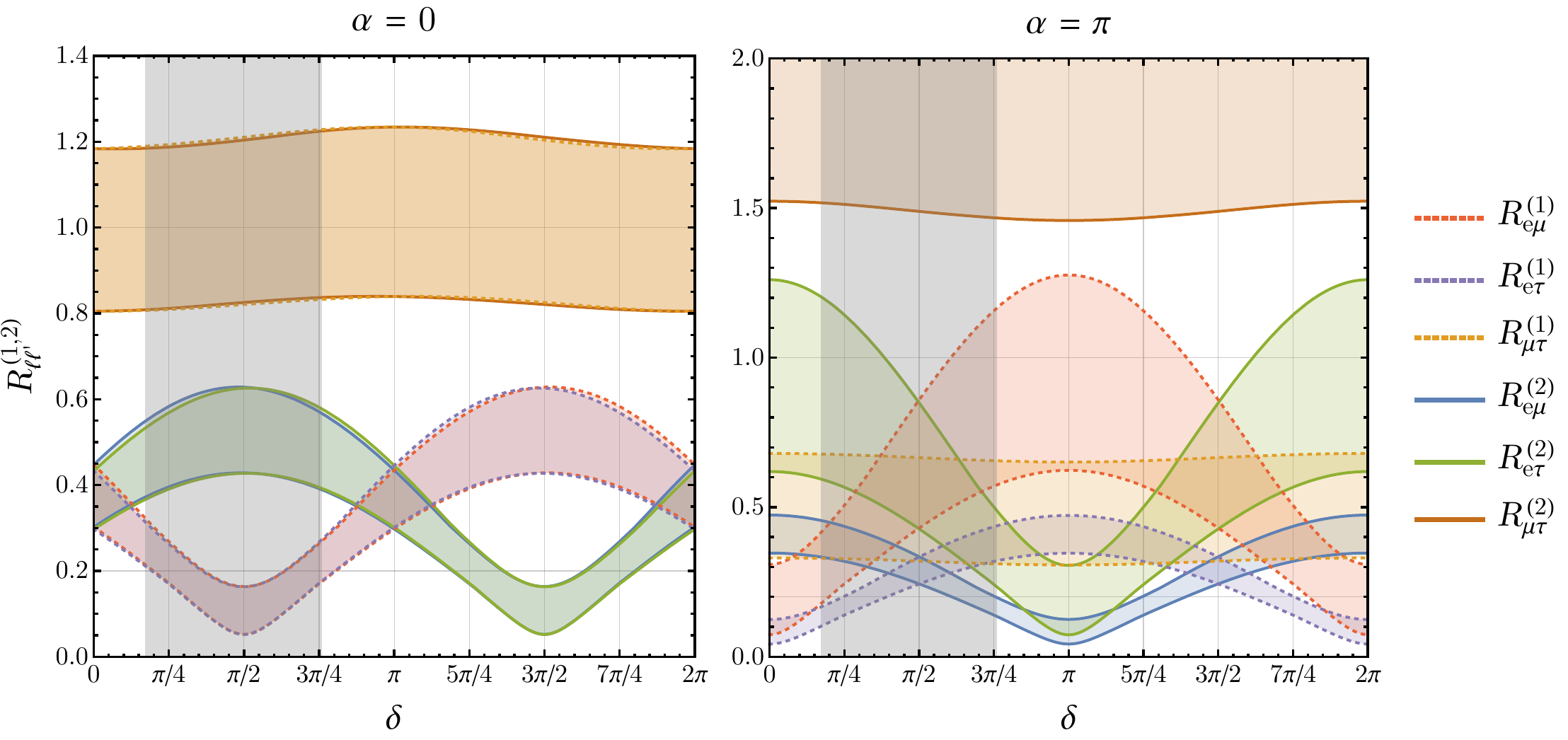}
\caption{Ratios $R^{(1,2)}_{\ell\ell'}$ of (absolute values of) Yukawa couplings
for a NO neutrino spectrum
as a function of the CPV phase $\delta$ for $\alpha = 0$ (left panel) and 
$\alpha = \pi$ (right panel),
in the case $\xi < 0$. Bands are obtained by varying $\Delta m^2_{21,31}$ and 
the $\sin^2 \theta_{ij}$ in their respective $3\sigma$ allowed ranges
given in Table \ref{tab:data}.
In the case $\alpha = \pi$, the upper boundary of the $R^{(2)}_{\mu\tau}$ band (not shown) is located at
$R^{(2)}_{\mu\tau} \simeq 3.0 - 3.2$.
The vertical grey band indicates values of $\delta$ which are disfavoured at $3\sigma$.
The case $\xi > 0$ is obtained by exchanging $R^{(1)}_{\ell\ell'}$ and $R^{(2)}_{\ell\ell'}$.}
\label{fig:ratios}
\end{figure}

 In Figs.~\ref{fig:ratiosbfv} and \ref{fig:ratios} we present the ratios
$R^{(1,2)}_{\ell\ell'}$ as a function of $\delta$
for the case $\xi < 0$ and two representative values of $\alpha$.
Figure \ref{fig:ratiosbfv} is obtained using the best fit values of
$\Delta m^2_{21,31}$ and $\sin^2 \theta_{ij}$ taken
from Table  \ref{tab:data}.   
In Fig.~\ref{fig:ratios} we show the ranges in which $R^{(1,2)}_{\ell\ell'}$ 
vary when $\Delta m^2_{21,31}$ and 
the $\sin^2 \theta_{ij}$ are varied in their respective
$3\sigma$ allowed intervals given in Table  \ref{tab:data}.
In Table \ref{tab:Rranges3s} we report the respective intervals in which
each of the six ratios can lie. As Table \ref{tab:Rranges3s} indicates,
certain specific simple patterns cannot be realised within the scheme
considered. Among those are, for example, the patterns 
$|g_{e 1}|:|g_{\mu 1}|:|g_{\tau 1}| \,\simeq\, 1:1:1$ and
$|g_{e 2}|:|g_{\mu 2}|:|g_{\tau 2}| \,\simeq\, 1:1:1$.
%%%%%%%%%%%%%%%%%%%%%%%%%%%%%%%
%%%%%%%%%%%%%%%%%%%%%%%%%%%%%%%
%%%%%%%%%%%%%%%%%%%%%%%%%%%%%%%
%%%%%%%%%%%%%%%%%%%%%%%%%%%%%%%
%
%%%%%%%%%%%%%%%%%%%%%%%%%%%%%%%
\begin{table}
\centering
\renewcommand{\arraystretch}{1.2}
\begin{tabular}{ c c }
  \hline			
  Ratio & Allowed range  \\ \hline			
  $R^{(1)}_{e\mu}$    & $0.05 - 1.28$  \\		
  $R^{(1)}_{e\tau}$   & $0.04 - 0.63$  \\		
  $R^{(1)}_{\mu\tau}$ & $0.31 - 1.23$ \\		
  $R^{(2)}_{e\mu}$    & $0.04 - 0.63$  \\		
  $R^{(2)}_{e\tau}$   & $0.05 - 1.26$  \\		
  $R^{(2)}_{\mu\tau}$ & $0.80 - 3.21$ \\		
  \hline  
\end{tabular}
\caption{Ranges for the ratios of absolute values of Yukawa couplings, obtained by 
varying $\Delta m^2_{21,31}$, the $\sin^2 \theta_{ij}$, and $\delta$
in their respective $3\sigma$ allowed ranges
and $\alpha$ in its defining range, for $\xi <0$.
The case $\xi > 0$ is obtained by exchanging $R^{(1)}_{\ell\ell'}$ and $R^{(2)}_{\ell\ell'}$.}
\label{tab:Rranges3s}
\end{table}
%%%%%%%%%%%%%%%%%%%%%%%%%%%%%%%
%

 The flavour structure of eq.~\eqref{eq:FNyukawas},
which is naturally realised in the model of Section~\ref{sec:model},
corresponds to the pattern
$|g_{e 2}| : |g_{\mu 2}| : |g_{\tau 2}| \,\simeq\, \epsilon : 1 : 1$,
and thus to 
$R^{(2)}_{e\mu} \simeq R^{(2)}_{e\tau} \simeq \epsilon$ 
and $R^{(2)}_{\mu\tau} \simeq 1$. 
The requirement of having $R^{(2)}_{\mu\tau} \simeq 1$ 
favours $\alpha$ close to zero~%
\footnote{%
Marginalizing over $\delta$ (either in its defining or in its $3\sigma$ range)
and varying $\Delta m^2_{21,31}$ and 
the $\sin^2 \theta_{ij}$ in their respective $3\sigma$ allowed ranges,
the requirement that $|R_{\mu\tau}^{(2)}-1| < 0.1$ implies 
$\alpha <0.36\pi \,\vee\,\alpha >1.64\pi$, independently of the sign of $\xi$.
However, if we require that the relative probability of $\alpha$
having a given value in the indicated intervals is not less than 0.15,
then we have $\alpha <0.2\pi$ or $\alpha >1.8\pi$.
For these values of $\alpha$ and $\epsilon = 0.2$,
the predictions for $\delta$ can be read off from the plots where $\alpha = 0$.
}.
As can be inferred from Fig.~\ref{fig:ratiosbfv},
given the current best fit values of neutrino mass
squared differences and mixing parameters,
the requirement of $R^{(2)}_{e\mu} \simeq R^{(2)}_{e\tau} \simeq \epsilon = 0.2$  
leads, for $\xi < 0$, to the prediction of
$\delta \simeq 5\pi/4,\,7\pi/4$~%
\footnote{Similar predictions for the $\delta$ and $\alpha$
were obtained in a different context in
Ref.~\cite{Nakayama:2017cij}.}.
Taking into account the $3\sigma$ allowed ranges of 
$\Delta m^2_{21,31}$ and $\sin^2 \theta_{ij}$ leads,
as Fig.~\ref{fig:ratios} shows,
to $\delta$ lying in narrow intervals around the values
$5\pi/4$ and $7\pi/4$.
Allowing for a somewhat smaller value of $\epsilon$, e.g.,
$\epsilon = 0.15$, we find that $\delta$ should lie in the interval
$\delta \simeq [5\pi/4,\,7\pi/4]$
which includes the value
$3\pi/2$
(see Fig.~\ref{fig:ratios}). 

 For $\delta \simeq 5\pi/4,\,7\pi/4$,
$\alpha = 0$ and
the best fit values of  $\Delta m^2_{21,31}$ and 
the $\sin^2 \theta_{ij}$ we get the following pattern of
the Yukawa couplings of $\nu_{1R}$:
$|g_{e 1}|:|g_{\mu 1}|:|g_{\tau 1}| \,\sim\, 0.5:1:1$.

 For $\xi > 0$, using the same arguments we obtain instead
$\delta \simeq \pi/4,3\pi/4$, or 
$\delta \simeq [\pi/4,3\pi/4]$. According to the
global analyses \cite{Capozzi:2017ipn,Esteban:2016qun},
however, these values of $\delta$ are strongly disfavoured
(if not ruled out) by the current data.

 In a more phenomenological approach, we get 
$\delta \simeq 3\pi/2$
provided, e.g., $|g_{e 2}| : |g_{\mu 2}| : |g_{\tau 2}| \,\simeq\, 0.14 : 1 : 1$
and $\alpha \simeq \pi/5$. In this case, the remaining ratios read
$|g_{e 1}| : |g_{\mu 1}| : |g_{\tau 1}| \,\simeq\, 0.5 : 0.7 : 1$.
In the GUT-inspired scenario of Ref.~\cite{Buchmuller:1998zf}, a different
FN charge assignment leads to $\epsilon = 0.06$, in which case
$\delta \simeq 3\pi/2$ is favoured.

%%%%%%%%%%%%%%%%%%%%%%% 
\section{Phenomenology}
\label{sec:pheno}
%%%%%%%%%%%%%%%%%%%%%%% 

 The low-energy phenomenology of the model of 
interest resembles that of the model 
with two heavy Majorana neutrinos $N_{1,2}$ forming a pseudo-Dirac pair 
considered in \cite{Ibarra:2010xw,Ibarra:2011xn,Dinh:2012bp}, 
in which the splitting between the masses of $N_{1,2}$ 
is exceedingly small.
For this model direct and indirect constraints
on the model's parameters, which do not depend on 
the splitting between the masses of $N_1$ and $N_2$,
as well as expected sensitivities of future lepton colliders
have been analysed, e.g.,
in Refs.~\cite{Ibarra:2010xw,Ibarra:2011xn,Dinh:2012bp,Antusch:2015mia,Das:2017nvm}
(see also~\cite{Deppisch:2015qwa,Das:2017zjc}).

 Due to the mixing of LH and RH neutrino fields,
i) the PMNS
neutrino mixing matrix, $\UPMNS$, as we have already noticed,
is not unitary, as also the expressions for the charged and neutral 
current weak interaction of the light Majorana neutrinos 
$\chi_i$ given in eqs.~\eqref{eq:CC0} and \eqref{eq:NC0} show, and
ii) the heavy Majorana neutrinos $N_{1,2}$ also participate
in charged and neutral current weak interactions
with the $W^\pm$ and $Z^0$ bosons:
%%%%%%%%%%%%%%%%%%%%%
\begin{align}
\mathcal{L}_\textrm{CC}^N \,&=\, - \frac{g}{\sqrt{2}}\,
\bar\ell\,\gamma_\alpha\,\big(RV\big)_{\ell k}\,N_{kL}\,W^\alpha
 \,+\, \textrm{h.c.} \,,\\
\mathcal{L}_\textrm{NC}^N \,&=\, - \frac{g}{2c_w}\,\overline{\nu_{\ell L}}\,\gamma_\alpha\,
\big(RV\big)_{\ell k}\,N_{kL}\,Z^\alpha \,+\, \textrm{h.c.}  \,.
\label{eq:CCNCheavy}
\end{align}
%%%%%%%%%%%%%%%%%%%%%
%
Due to the Yukawa interactions, cf.~eq.~\eqref{eq:Yukawa},
there are interactions of the heavy Majorana neutrinos $N_{1,2}$
with the SM Higgs boson $h$ as well (see \cite{Cely:2012bz}):
%%%%%%%%%%%%%%%%%%%%%
\begin{equation}
\mathcal{L}_\textrm{H}^N \,=\, -\,\frac{M_k}{v}\,
\overline{\nu_{\ell L}}\, \big(RV\big)_{\ell k} \, N_{kR}\, h
 \,+\, \textrm{h.c.} \,.
\label{eq:intHiggs}
\end{equation}
%%%%%%%%%%%%%%%%%%%%%

%%%%%%%%%%%%%%%%%%%%%%%%%%%%
\subsection{Neutrino mass matrix and non-unitarity bounds}
\label{subsection:massn}
%%%%%%%%%%%%%%%%%%%%%%%%%%%%

 The first constraint on the $RV$ elements follows from the fact 
that the elements of the light neutrino Majorana mass matrix, 
$(m_{\nu})_{\ell \ell'}$, have rather small maximal values. 
Indeed, as it follows from eq.~\eqref{eq:mnu0}, 
we have \cite{Ibarra:2010xw}:
%
%%%%%%%%%%%%%%%%%%%%%%%%%%%%%%%
\begin{equation}
|(m_{\nu})_{\ell \ell'}| = |U^*_{\ell j}\, m_j\, U^*_{\ell' j}|  
\simeq \left |\sum_{a} (RV)^*_{\ell a}\, M_a\, (RV)^*_{\ell' a} \right |
\,,~\ell,\ell'=e,\mu,\tau\,,
\label{eq:mnuRV}
\end{equation}
%%%%%%%%%%%%%%%%%%%%%%%%%%%%%%%
%
where the sum is effectively over $j=2,3$ since in the model considered 
$m_1 = 0$ 
\footnote{Strictly speaking, we have $m_1 = 0$ only at tree level.
Higher order corrections lead to  a non-zero value of $m_1$, which
is however negligibly small.
}. 
The elements of the neutrino Majorana mass matrix
$(m_{\nu})_{\ell \ell'}$ depend, apart from  
$m_2 = \sqrt{\Delta m^2_{21}} \simeq  8.6\times 10^{-3}$ eV, 
$m_3 = \sqrt{\Delta m^2_{31}}\simeq 0.051$ eV,
$\theta_{12}$, $\theta_{23}$, $\theta_{13}$, 
on the CPV phases $\delta$ and $\alpha$. 
The maximal value a given element of  $m_{\nu}$ can have 
depends on its flavour indices $\ell$ and $\ell'$.
It is not difficult to derive these maximal values using the results 
reported in Table \ref{tab:data}.
We have:
\begin{enumerate}[i)]
\item $|(m_{\nu})_{e e }|       \ltap 4.3\times 10^{-3}$ eV ($\alpha + 2\delta = 0$);
\item $|(m_{\nu})_{e \mu }|     \ltap 9.2\times 10^{-3}$ eV ($\delta = \pi,\,\alpha = \pi$);
\item $|(m_{\nu})_{e \tau }|    \ltap 9.2\times 10^{-3}$ eV ($\delta = 0  ,\,\alpha = \pi$);
\item $|(m_{\nu})_{\mu \mu }|   \ltap 3.4\times 10^{-2}$ eV ($\delta = \pi,\,\alpha = 0  $);
\item $|(m_{\nu})_{\mu \tau }|  \ltap 2.9\times 10^{-2}$ eV ($\delta = 3\pi/2,\,\alpha = \pi$);
\item $|(m_{\nu})_{\tau \tau }| \ltap 3.5\times 10^{-2}$ eV ($\delta = 0  ,\,\alpha = 0$).
\end{enumerate}
The quoted maximal values are reached for the values of the CPV phases 
given in the brackets. It should be added that the dependence of 
${\rm max}(|(m_{\nu})_{\ell \ell' }|)$, $\ell,\ell'=\mu,\tau$, on $\delta$
and $\alpha$ is rather weak since the terms involving $\delta$ always include 
the suppressing factor $\sin\theta_{13}$, while the term $\propto m_2$ 
is considerably smaller (typically by a factor of 10) than 
the term  $\propto m_3$ as $m_2/m_3 \simeq 0.17$.
We will consider $|(m_{\nu})_{e e }| \ltap 4\times 10^{-3}$ eV,
$|(m_{\nu})_{e \mu}|,|(m_{\nu})_{e \tau}| \ltap 9\times 10^{-3}$ eV,
and $|(m_{\nu})_{\ell \ell'}| \ltap 3\times 10^{-2}$ eV, 
$\ell,\ell'=\mu,\tau$,
as reference maximal values in the numerical 
analysis which follows.

 From the expression for $RV$ given in eq.~\eqref{eq:RVgeneral}
and eq.~\eqref{eq:mnuRV}, and taking into account the mass splitting
between $N_1$ and $N_2$, we get to leading order in 
$|g_{\ell 1}|$, $|g_{\ell' 2}|$ and $|g_{\ell 1}g_{\ell' 2}|$:
%%%%%%%%%%%%%%%%%%%%%%%%%%%%%
\begin{equation}
\left |(m_{\nu})_{\ell \ell'} \right | \,\simeq\, 
\frac{v^2}{M}\,
\left | g_{\ell 1}g_{\ell' 2} +  g_{\ell 2}g_{\ell' 1} \right | + 
\mathcal{O}(g_{\ell 1}g_{\ell' 1}) \,,
\label{eq:mnug1g2}
\end{equation}
%%%%%%%%%%%%%%%%%%%%%%%%%%%%%
%
which coincides (up to higher order corrections) 
with the form given in eq.~\eqref{eq:mnu}.
Thus, for a given value of $M$,
the upper bounds on  $|(m_{\nu})_{\ell \ell'}|$ 
lead via  eq.~\eqref{eq:mnuRV} 
to upper bounds on the magnitude of the product of 
the neutrino Yukawa couplings of $\nu_{1R}$ and   $\nu_{2R}$,  
$g_{\ell 1}$ and $g_{\ell' 2}$. As we have seen, 
these bounds depend on the flavour of the lepton doublet 
to which  $\nu_{1R}$ and $\nu_{2R}$ are coupled. 

 For $M = 100$ GeV (1 TeV), for example,
the constraint of interest $|(m_{\nu})_{e e}| \ltap 4\times 10^{-3}~{\rm eV}$ implies 
$2|g_{e 1}g_{e 2}| \ltap 1.3\times 10^{-14}$ ($1.3\times 10^{-13}$). 
This upper limit can be satisfied
for, e.g., $|g_{e 1}| \sim 0.65\times 10^{-12}$ ($0.65\times 10^{-11}$) and  
$|g_{\ell' 2}| \sim 10^{-2}$. The upper bounds on
$|g_{e1}g_{\ell 2} +  g_{\ell 1}g_{e 2}|$, $\ell=\mu,\tau$, is approximately
by a factor of 2 larger than the quoted upper bound on
$2|g_{e 1}g_{e 2}|$, while those on 
$| g_{\ell 1}g_{\ell' 2} +  g_{\ell 2}g_{\ell' 1}|$, $\ell,\ell'=\mu,\tau$
are larger approximately by a factor of 8.

 In \cite{Ibarra:2010xw,Ibarra:2011xn} the constraint in 
eq.~\eqref{eq:mnuRV} is satisfied by finding a region, in the general 
parameter space of the model considered, in which to leading order 
$\sum_{a=1,2}\,(RV)^*_{\ell a}\, M_a\, (RV)^*_{\ell' a}$ = 0, i.e., 
the two terms in the sum cancel. 
In the version of the low-scale type I seesaw model 
with two RH neutrinos we are considering the 
constraint in eq.~\eqref{eq:mnuRV}
is satisfied due to smallness of the product of Yukawa 
couplings $|g_{\ell 1}|$ and $|g_{\ell' 2}|$. 
In the model under consideration one gets 
$\sum_{a=1,2}\,(RV)^*_{\ell a}\, M_a\, (RV)^*_{\ell' a}$ = 0 
in the limit of negligible couplings $g_{\ell 1}$.
Indeed, setting $g_{\ell 1} = 0$ we get $M_1 = M_2$ 
and the expression for the matrix RV takes the form:
%%%%%%%%%%%%%%%%%%%%%
\begin{equation}
RV \,\simeq\, 
\frac{1}{\sqrt{2}}\,
\frac{v}{M}\,
\left(\begin{array}{cc}
g_{e2}^*     & -i\,g_{e2}^* \\
g_{\mu 2}^*   & -i\,g_{\mu 2}^* \\
g_{\tau 2}^*  & -i\,g_{\tau 2}^* 
\end{array}\right)\,.
\label{eq:RVg2}
\end{equation}
%%%%%%%%%%%%%%%%%%%%%%%%%%%%%%%%
%
This implies
%%%%%%%%%%%%%%%%%%%%%%%%%%%%%%
\begin{equation}
(RV)_{\ell 1} = -i\, (RV)_{\ell 2}\,,~~l=e,\mu,\tau\,,
\label{eq:RV1RV2}
\end{equation}
%%%%%%%%%%%%%%%%%%%%%%%%%%%%%%%
%
which together with the equality $M_1 = M_2$ leads 
\footnote{The same relation \eqref{eq:RV1RV2}
holds in the limit of zero splitting between the masses 
of $N_1$ and $N_2$ in the version of the TeV scale 
type I seesaw model considered 
in \cite{Ibarra:2011xn,Dinh:2012bp}.}
to $\sum_{a=1,2}\,(RV)^*_{\ell a}\, M_a\, (RV)^*_{\ell' a}$ = 0.

 As we have already discussed, the matrix 
$\eta \equiv - R\, R^\dagger/2 =  - (RV)\,(RV)^\dagger/2
= \eta^\dagger$ parametrises the
deviations from unitary of the PMNS matrix.
The elements of $\eta$ are constrained by precision electroweak data 
and data on flavour observables. 
For heavy Majorana neutrino masses above the electroweak scale
the most updated set of constraints on the absolute values of 
the elements of $\eta$ at $2\sigma$ C.L. 
reads~\cite{Fernandez-Martinez:2016lgt,Blennow:2016jkn}:
%%%%%%%%%%%%%%%%%%%%%
\begin{equation}
|\eta| \,<\,  
\left(
\begin{array}{ccc}
 1.3 \times 10^{-3} & 1.2 \times 10^{-5} & 1.4 \times 10^{-3} \\
 1.2 \times 10^{-5}  & 2.2 \times 10^{-4} & 6.0 \times 10^{-4} \\
 1.4 \times 10^{-3} & 6.0 \times 10^{-4} & 2.8 \times 10^{-3} \\
\end{array}
\right)\,.
\label{eq:NU}
\end{equation}
%%%%%%%%%%%%%%%%%%%%%
%
The upper bound on the $e-\mu$ elements is relaxed to
$|\eta_{e\mu}| <  3.4 \times 10^{-4}$
for heavy Majorana neutrino masses below the electroweak scale
(but still above the kaon mass, $M_k \gtap 500$ MeV)
due to the restoration of a GIM cancellation 
\cite{Petcov:1976ff}. The above constraints on $\eta$
justify the assumption made in Section~\ref{sec:setup}
regarding the smallness of the elements of $R$.

 Using the expression for $RV$ given in eq.~\eqref{eq:RVgeneral}
we find that, to leading order in  $g_{\ell 1}$, 
$g_{\ell' 2}$, $|g_{\ell 1}| \ll |g_{\ell' 2}|$, we have:
%%%%%%%%%%%%%%%%%%%%%%%%%%%%%%%
\begin{equation}
|\eta_{\ell \ell'}| \simeq \frac{1}{2}\,\frac{v^2}{M^2}\,
\left | g_{\ell 2}\,g_{\ell' 2} \right | + 
\mathcal{O}(g_{\ell 1}\,g_{\ell' 2},g_{\ell' 1}\,g_{\ell 2})\,.
\label{eq:etag2g}
\end{equation}
%%%%%%%%%%%%%%%%%%%%%%%%%%%%%%
%
As a consequence, if $M$ is given, 
the experimental limits on $|\eta|$ cited in eq.~\eqref{eq:NU}, 
in contrast to the limits on $|(m_{\nu})_{\ell \ell'}|$, 
imply upper bounds on $|g_{\ell 2}\,g_{\ell' 2}|$, i.e., 
on the Yukawa couplings of $\nu_{2R}$. 
For, e.g., $M=100$ GeV we find, depending on the flavour indices,  
$|g_{\ell 2}\, g_{\ell' 2}|^{1/2} 
\ltap (2.8\times 10^{-3} - 4.3\times 10^{-2})$,
i.e., $|g_{\ell 2}|$ can be relatively large. 
This can lead to interesting low-energy phenomenology 
involving the heavy Majorana neutrinos $N_{1,2}$.

%%%%%%%%%%%%%%%%%%%%%%%%%%%
\subsection{LFV Observables and Higgs Decays}
\label{subsection:LFV}
%%%%%%%%%%%%%%%%%%%%%%%%%%%%

 The predictions of the model under discussion for the 
rates of the lepton flavour violating (LFV) 
$\mu \rightarrow e\gamma$ and $\mu \rightarrow eee$ decays and 
$\mu -  e$ conversion in nuclei, as can be shown, depend on 
$|(RV)_{\mu 1}^* (RV)_{e 1} + (RV)_{\mu 2}^* (RV)_{e 2} |^2 \simeq 
4\,|(RV)_{\mu 2}^* (RV)_{e 2} |^2$, where we have used eq.~\eqref{eq:RV1RV2}, 
and on the masses $M_1 \simeq M_2 \simeq M$ of 
the heavy Majorana neutrinos $N_1$ and $N_2$. 
The expressions for the $\mu \rightarrow e\gamma$ and 
$\mu \rightarrow eee$ decay branching ratios, 
$\textrm{BR}(\mu \rightarrow e \gamma)$ and 
$\textrm{BR}(\mu \rightarrow eee)$, and for the relative 
$\mu -  e$ conversion in a nucleus $X$, 
$\textrm{CR}(\mu X \rightarrow eX)$, coincide with those given in 
Refs.~\cite{Ibarra:2011xn,Dinh:2012bp} and we are not going 
to reproduce them here. The best experimental limits on 
$\textrm{BR}(\mu \rightarrow e \gamma)$,
$\textrm{BR}(\mu \rightarrow eee)$ and $\textrm{CR}(\mu X \rightarrow eX)$ 
have been obtained by the MEG \cite{TheMEG:2016wtm}, 
SINDRUM \cite{Bellgardt:1987du}
and SINDRUM II \cite{Dohmen:1993mp,Bertl:2006up} Collaborations:
%%%%%%%%%%%%%%%%%%%%%%%%%
\begin{align}
\textrm{BR}(\mu \rightarrow e\gamma) \,&<\, 4.2 \times 10^{-13}~\textrm{(90\% C.L.)}\,,
\label{eq:MEG} \\
\textrm{BR}(\mu \rightarrow eee) \,&<\, 1.0 \times 10^{-12}~\textrm{(90\% C.L.)}\,,
\label{eq:SINDRUM} \\
\textrm{CR}(\mu\,\textrm{Ti} \rightarrow e\,\textrm{Ti}) \,&<\, 4.3 \times 10^{-12}~\textrm{(90\% C.L.)}\,,
\label{eq:SINDRUMII} \\
\textrm{CR}(\mu\,\textrm{Au} \rightarrow e\,\textrm{Au}) \,&<\, 7 \times 10^{-13}~\textrm{(90\% C.L.)}\,.
\label{eq:SINDRUMII2}
\end{align}
%%%%%%%%%%%%%%%%%%%%%
%

 The planned MEG II update of the MEG experiment~\cite{Cattaneo:2017psr} 
is expected to reach sensitivity to
$\textrm{BR}(\mu \rightarrow e\gamma) \simeq 4 \times 10^{-14}$.
The sensitivity to $\textrm{BR}(\mu \rightarrow eee)$  
is expected to experience a dramatic increase of up to four orders 
of magnitude with the realisation of the Mu3e Project~\cite{Blondel:2013ia}, 
which aims at probing values down to 
$\textrm{BR}(\mu \rightarrow eee) \,\sim\, 10^{-16}$ in its phase II 
of operation.
Using an aluminium target, the Mu2e~\cite{Bartoszek:2014mya} and 
COMET~\cite{Kuno:2013mha} collaborations plan to ultimately be sensitive to 
$\textrm{CR}(\mu\,\textrm{Al} \rightarrow e\,\textrm{Al}) \sim\, 
6 \times 10^{-17}$. The PRISM/PRIME project~\cite{Barlow:2011zza} 
aims at an impressive increase of sensitivity to 
the $\mu - e$ conversion rate in titanium, planning to probe values down to
$\textrm{CR}(\mu\,\textrm{Ti} \rightarrow e\,\textrm{Ti}) \,\sim\, 10^{-18}$,
an improvement of six orders of magnitude with respect to the bound 
of eq.~\eqref{eq:SINDRUMII}. 
%%%%%%%%%%%%%%%%%%%%%%%%%%%%
\begin{figure}[t]
\centering
\includegraphics[width=\textwidth]{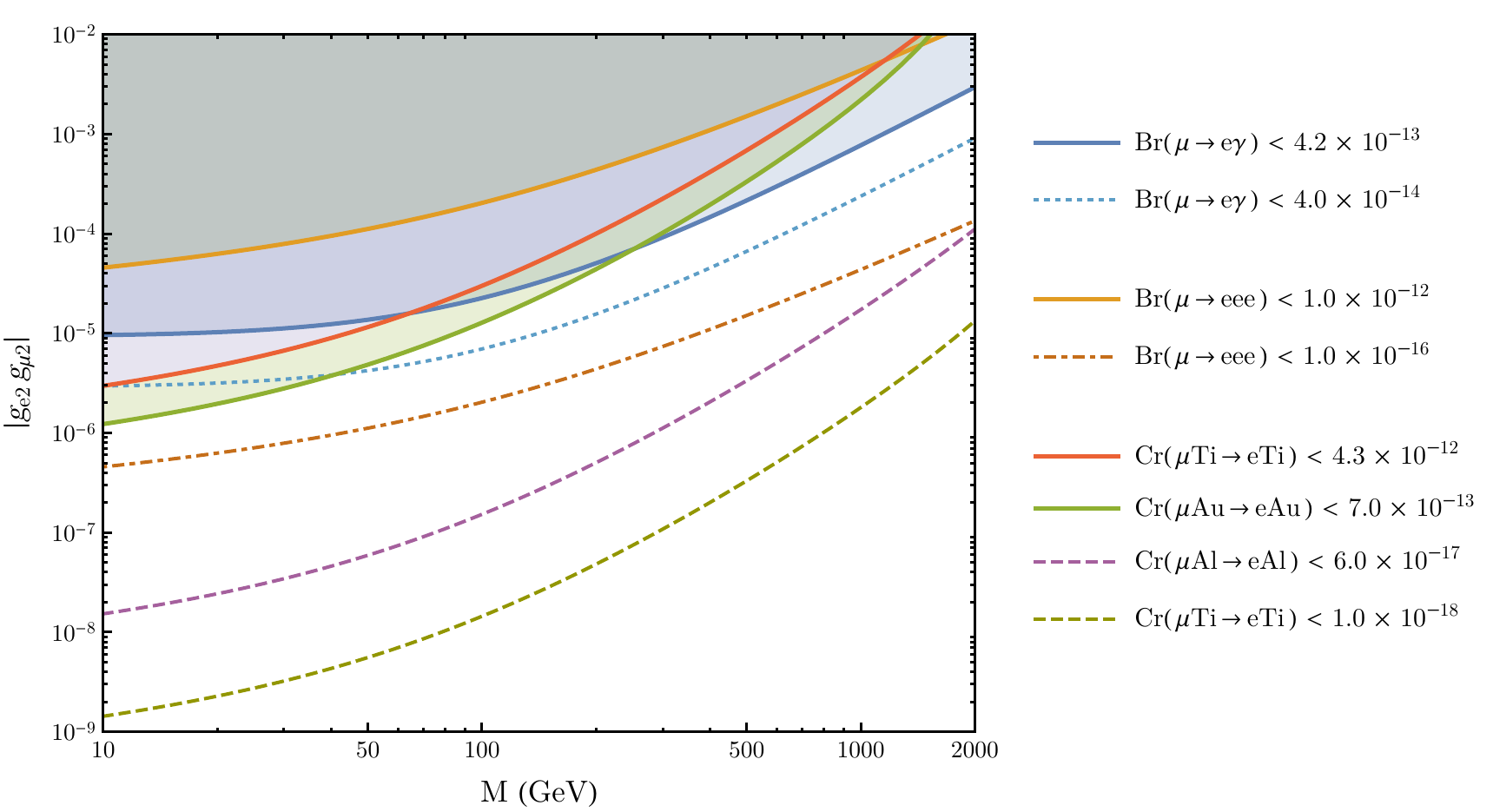}
\caption{Present limits (solid lines) and expected future sensitivities 
(dotted, dashed and dot-dashed lines) on $|g_{\mu 2}||g_{e 2}|$ 
from data on muon LFV processes,   
as a function of the mass $M$ 
of heavy Majorana neutrinos. See text for details.}
\label{fig:LFVemu}
\end{figure}
%%%%%%%%%%%%%%%%%%%%%%%%%%%%%%%%

 We show in Fig.~\ref{fig:LFVemu} the limits on $|g_{\mu 2}\,g_{e 2}|$
implied by the experimental bounds in 
eqs.~\eqref{eq:MEG}~--~\eqref{eq:SINDRUMII2}, 
as a function of the mass $M$, as well as the prospective 
sensitivity of the future planned experiments 
MEG II, Mu3e,  Mu2e, COMET and  PRISM/PRIME.
The data from these experiments, as Fig.~\ref{fig:LFVemu} 
indicates, will allow to test for values of 
 $|g_{\mu 2}\,g_{e 2}|$ significantly smaller 
than the existing limits, with a significant potential 
for a discovery. 

 The interactions given in eq.~\eqref{eq:intHiggs} 
open up novel decay channels
for the Higgs boson, provided the masses of the
heavy neutrinos $N_{1,2}$ are below the Higgs boson mass.
For $M_{1,2} < m_h = 125.1\,\GEV$, the new Higgs decay modes
are those into one light and one heavy neutrino, 
$h\rightarrow \nu_{\ell L}\, N_{k}$, $\ell=e,\mu,\tau$, $k=1,2$.
The phenomenology of the Higgs decays $h\rightarrow \nu_{\ell L}\, N_{k}$ 
in the model considered in the present article
is similar to that of the same decay investigated in 
detail in \cite{Cely:2012bz} 
in the model discussed in \cite{Ibarra:2011xn}.
The rate of the decay $h\rightarrow \nu_{\ell L}\, N_{1,2}$ 
to any $\nu_{\ell L}$ and $N_{1}$ or $N_2$ is given in Ref.~\cite{Cely:2012bz} 
and in the limit of zero mass splitting of  $N_{1,2}$ 
($M_1 = M_2 = M$) reads:
%%%%%%%%%%%%%%%%%%%%%
\begin{equation}
\Gamma(h \rightarrow \nu\, N) \,=\,
\frac{m_h}{16\pi}  
\,\bigg(1 - \frac{M^2}{m_h^2} \bigg)^2
\,  \frac{M^2}{v^2}
\,
\sum_{\ell,k}\big|\big(RV\big)_{\ell k}\big|^2
\,,
\label{eq:decHiggs}
\end{equation}
%%%%%%%%%%%%%%%%%%%%%
%
where in the model considered by us 
%%%%%%%%%%%%%%%%%%%%%%%%%%%%
\begin{equation}
  \frac{M^2}{v^2}
\,\sum_{\ell,k}\big|\big(RV\big)_{\ell k}\big|^2 = 
|g_{e 2}|^2 + |g_{\mu 2}|^2 + |g_{\tau 2}|^2 \,,
\label{eq:RVHiggs}
\end{equation}
%%%%%%%%%%%%%%%%%%%%%%%%%%%
%
and we have used eqs.~\eqref{eq:RVg2} and \eqref{eq:RV1RV2}.
The dominant decay mode of the SM Higgs boson is into 
bottom quark-antiquark pair, $b-\bar{b}$. The decay rate is given by: 
%%%%%%%%%%%%%%%%%%%%%%%%%%%%
\begin{align}
\Gamma (h \rightarrow b\,\bar{b}) = \frac{3m_h}{16\pi}\left(\frac{m_b}{v}\right)^2\,
\left(1 - \frac{4m^2_b}{m^2_h}\right)^{3/2}\,,
\end{align}
%%%%%%%%%%%%%%%%%%%%%%%%%%%%
%
$m_b \simeq 4.18$ GeV being 
the $b-$quark mass (in the $\overline{\text{MS}}$ scheme). The SM branching ratio of this decay 
is 58.4\% \cite{Olive:2016xmw2}. The total SM decay width 
of the Higgs boson is rather small \cite{Olive:2016xmw2}:
$\Gamma^{\rm SM}_{\rm tot} \simeq 4.07\times 10^{-3}$ GeV.

 The upper bound on  $(\sum_{\ell} |g_{\ell 2}|^2)$ is determined 
essentially by the upper bound on  
$|g_{\tau 2}|^2 = 2|\eta_{\tau \tau}|M^2/v^2$, which is less 
stringent than the upper bounds on $|g_{e 2}|^2$ and  
$|g_{\mu 2}|^2$. Using the bound 
$|\eta_{\tau \tau}| < 2.8\times 10^{-3}$ quoted in eq.~\eqref{eq:NU}, 
we get for $M = 100$ GeV the upper bound 
$|g_{\tau 2}|^2 < 1.8\times 10^{-3}$.
For the Higgs decay rate 
$\Gamma(h \rightarrow \nu\, N)$ in the case of 
$M = 100$ GeV and, e.g.,  
$(\sum_{\ell} |g_{\ell 2}|^2) = 10^{-3}$,
we get $\Gamma(h \rightarrow \nu\, N) = 3.2 \times10^{-4}$ 
GeV. This decay rate 
would lead to an increase of the total SM decay width of 
the Higgs boson by approximately 8\%.
Thus, the presence of the  $h \rightarrow \nu\, N$ decay would modify 
the SM prediction for the branching ratio for any generic 
(allowed in the SM) decay of the Higgs particle \cite{Cely:2012bz}, 
decreasing it.  

 We finally comment on neutrinoless double 
beta ($(\beta\beta)_{0\nu}$-) decay 
(see, e.g., \cite{Olive:2016xmw}).
The relevant observable is the absolute value of the effective neutrino 
Majorana mass $|\langle m \rangle|$ (see, e.g., \cite{Bilenky:1987ty}), 
which receives an extra contribution from the exchange 
of heavy Majorana neutrinos $N_1$ and $N_2$. This contribution 
should be added to that due to the light Majorana neutrino exchange 
\cite{Halprin:1983ez,Haxton:1985am} (see also 
\cite{Ibarra:2010xw,Lopez-Pavon:2015cga}).
The sum of the two contributions can lead, in principle, to 
$|\langle m \rangle|$ that differs significantly 
from that due to the light Majorana neutrino exchange.
The contribution due to the $N_{1,2}$ exchange in 
$|\langle m \rangle|$ in the model considered is proportional, 
in particular, to the difference 
between the masses of $N_1$ and $N_2$, which form a 
pseudo-Dirac pair. For $M \gtap 1$ GeV, as can be shown, 
it is strongly  suppressed in the present setup due to the
extremely small $N_1 - N_2$ mass difference, 
the stringent upper limit on $|g_{e2}|^2$, 
and the values of the relevant nuclear matrix elements (NME), 
which at $M = 1$ GeV are smaller approximately by a factor of 
$6\times 10^{-2}$ than the NME for the light neutrino exchange 
and scale with $M$ as $(0.9~\text{GeV}/M)^2$. 
As a consequence, the contribution to $|\langle m \rangle|$ 
due to the exchange of  $N_1$ and $N_2$ is 
significantly smaller than the contribution from the exchange of 
light Majorana neutrinos $\chi_j$.

%%%%%%%%%%%%%%%%%%%%%%%%%%%%%%%%
\section{Summary and Conclusions}
\label{sec:conclusions}
%%%%%%%%%%%%%%%%%%%%%%%%%%%%%%%%

 In the present paper we have explored a symmetry-protected
scenario of neutrino mass generation, where two RH neutrinos 
are added to the SM. 
In the class of models considered, the main source of 
$L$-violation responsible for the neutrino masses
are small lepton-charge violating Yukawa couplings 
$g_{\ell 1}\,(\ell = e,\mu,\tau)$ to one of the RH neutrinos, $\nu_{1R}$.
Thus, the smallness of the light Majorana neutrino masses 
is related to the smallness of the $g_{\ell 1}$ and not to the 
RH neutrinos having large Majorana masses in the range of 
$\sim (10^{10} - 10^{14})$ GeV as in the standard seesaw scenario.
We have considered heavy Majorana neutrinos 
forming a pseudo-Dirac pair with masses $M_{1,2} \simeq M$ at the TeV or lower scale,
which are potentially observable in collider experiments.

 The setup described above can be realised in a Froggatt-Nielsen (FN) scheme, 
as detailed in Section \ref{sec:model}. In such a model, no U(1)$_L$ symmetry is imposed,
and instead the suppression of $L$-violating operators arises in the limit
of a large FN charge for $\nu_{1R}$. 
The FN charge assignments are partly motivated by
large $\nu_\mu$ -- $\nu_\tau$ mixing.
The structure of the Yukawa couplings $g_{\ell a}\,(a = 1,2)$ is
then determined by the FN charges, and yields 
$|g_{e 2}| : |g_{\mu 2}| : |g_{\tau 2}| \,\simeq\, \epsilon : 1 : 1$,
where $\epsilon \simeq \lambda_C \simeq 0.2$ is the FN suppression parameter,
while no unambiguous prediction may be extracted for the ratios
$|g_{e 1}| : |g_{\mu 1}| : |g_{\tau 1}|$.

 It is interesting to point out that,
given the exceedingly small splitting between heavy neutrinos,
the dependence on the Casas-Ibarra complex parameter drops out in
the ratios between absolute values of Yukawa couplings
to the same RH neutrino.
These ratios are then determined
(up to the exchange of $g_{\ell 1}$ and $g_{\ell 2}$)
by neutrino low-energy parameters alone,
namely, by neutrino masses, mixing angles
and CPV phases $\delta$ and $\alpha$.
Given the Yukawa structure of our model,
$|g_{e 2}| : |g_{\mu 2}| : |g_{\tau 2}| \,\simeq\, \epsilon : 1 : 1$ 
with $\epsilon \simeq \lambda_C \simeq 0.2$,
the Dirac CPV phase $\delta$ is predicted to have 
approximately one of the values 
$\delta \simeq \pi/4,\, 3\pi/4,$ or
$5\pi/4,\, 7\pi/4$,
or to lie in a narrow interval around one of these values,  
while a Majorana CPV phase $\alpha \simeq 0$ is preferred 
(Figs.~\ref{fig:ratiosbfv} and \ref{fig:ratios}).

 In the considered scenario, the maximal values of the elements
of the neutrino mass matrix lead to constraints on the combinations
$|g_{\ell 1}g_{\ell' 2}+g_{\ell' 1}g_{\ell 2}|$, $\ell,\ell'=e,\mu,\tau$,
which depend on products of $L$-conserving and $L$-violating Yukawa couplings
(see Section \ref{subsection:massn}).
Deviations from unitarity of the PMNS matrix constrain instead the products
$|g_{\ell 2}g_{\ell' 2}|$, $\ell,\ell'=e,\mu,\tau$, of $L$-conserving couplings alone.
In particular, the product $|g_{\mu 2}g_{e 2}|$ is constrained
by data on muon lepton flavour violating (LFV) processes.
Data from future LFV experiments (MEG II, Mu3e,  Mu2e, COMET, PRISM/PRIME)
will allow to probe values of $|g_{\mu 2}\,g_{e 2}|$
significantly smaller than the existing limits (Fig.~\ref{fig:LFVemu}).
The decay of the Higgs boson into one light and one heavy neutrino
can have a rate $\Gamma(h \rightarrow \nu N)$ as large as 8\% of the
total SM Higgs decay width. This decay mode can lead to a change of the Higgs branching
ratios with respect to the SM predictions.
Concerning neutrinoless double beta decay in the considered model, 
the contribution due to $N_{1,2}$ exchange in
the absolute value of the effective neutrino 
Majorana mass $|\langle m \rangle|$ is found to be negligible
when compared to the contribution from the exchange of 
light Majorana neutrinos.

 Finally, we comment on the issue of leptogenesis.
For temperatures above the electroweak phase transition (EWPT),
the Higgs VEV vanishes and thus, in the considered setup,
the splitting between the masses of heavy neutrinos originates
from the (suppressed) Majorana mass term
$\mu\, \nu^T_{1R}\,C^{-1}\,\nu_{1R}$,
with $\mu \sim \epsilon^{n+1} M \sim |g_{\ell 1}| M$. 
This component of the heavy neutrino mass matrix --
which in our case presents a subleading 
contribution to neutrino masses --
is then crucial for resonant leptogenesis 
to proceed (see, e.g.,~\cite{Dev:2014laa}).
The resonant condition reads $\mu \simeq \Gamma / 2$,
where $\Gamma$ denotes the average heavy neutrino decay width.
However, the values of $\mu$, $\Gamma$ and neutrino masses are tightly connected
in the FN model we analyse, which, together with the required 
smallness of $\mu$, prevents reproducing the observed 
baryon asymmetry of the Universe (BAU),
$\eta_B^\text{obs} \simeq (6.09 \pm 0.06) \times 10^{-10}$ \cite{Ade:2015xua}.

 One may instead successfully generate the observed BAU
through the mechanism of anti-leptogenesis \cite{Fukugita:2002hu}
(also known as ``neutrino assisted GUT baryogenesis'').
In this case, an excess of both baryon number $B$ and lepton number
$\hat{L}$ (see Section \ref{sec:model})
is produced at a high energy scale
($T > 10^{12}$ GeV, possibly related to grand unification),
while conserving $B-\hat{L}$. 
If there are new $\hat{L}$-violating interactions 
in thermal equilibrium at
such high temperature, they may erase the lepton number excess while leaving
the baryon number excess untouched, since sphalerons are not efficient
at these times.
At later times, sphalerons are responsible for only a partial conversion
of the baryon number excess into a lepton number excess, 
while some of the baryon excess remains.
Unlike resonant leptogenesis, this mechanism relies 
on a suppression of the $\hat{L}$-violating heavy neutrino 
mass splitting above the EWPT,
in order not to wash\discretionary{-}{-}{-}out the asymmetry 
generated at a high scale.
Modifying our setup as detailed in the end of Section \ref{sec:model},
the Majorana mass term $\mu\, \nu^T_{1R}\,C^{-1}\,\nu_{1R}$
is forbidden and the heavy neutrinos are degenerate above the EWPT.
One then adds a third RH neutrino in the bulk
with $(B-\hat{L})(\nu_{3R}) = - 1$ and vanishing U(1)$_{L}$ charge, 
such that its Yukawa couplings, 
which violate lepton number,
are allowed,
and such that the mass term
$M_3\, \nu^T_{3R}\,C^{-1}\,\nu_{3R}$ 
is generated, $M_3 \sim \langle \Phi\rangle$.
Notice that only one such RH neutrino
is needed to erase lepton number at high temperatures
($M_3 \sim (10^{12} - 10^{13})$ GeV),
and that there is a large region of parameter space where
the new contribution to the neutrino mass matrix is negligible 
\cite{Huang:2016wwj}.
Given these conditions, successful anti-leptogenesis may proceed.

%%%%%%%%%%%%%%%%%%%%%%%%
\section*{Acknowledgements} 
%%%%%%%%%%%%%%%%%%%%%%%%

 J.T.P.~would like to thank the Kavli IPMU, 
where part of the work on the present article was done, for kind hospitality.
This work was supported in part by the INFN
program on Theoretical Astroparticle Physics (TASP), 
by the European Union Horizon 2020 research and innovation programme
under the  Marie Sklodowska-Curie grants 674896 and 690575 (J.T.P.~and S.T.P.),
by Grant-in-Aid for Scientific Research from the Ministry of Education,
Science, Sports, and Culture (MEXT), Japan, No.~26104009, No.~26287039 and 
No.~16H02176 (T.T.Y.), 
and by the World Premier International Research Center Initiative (WPI
Initiative), MEXT, Japan (S.T.P. and T.T.Y.).

%%%%%%%%%%%%%%%%%%%%%%% Bibliography

\end{document}